\documentclass[aip,pop,numerical,preprint]{revtex4-1} 

\usepackage{graphicx}
\usepackage{amsmath,amssymb}
\usepackage{bm}

\newcommand{\Eq}{Equation~}
\newcommand{\Eqs}{Equations~}
\newcommand{\eq}{Equation~}
\newcommand{\eqs}{Equations~}

\newcommand{\fig}{Figure~}

\newcommand{\ie}{i.e.,}

\newcommand{\dd}{{\rm d}}
\newcommand{\ee}{{\rm e}}
\newcommand{\ii}{{\rm i}}
\newcommand{\nablab}{{\bm{\nabla}}}
\newcommand{\sech}{{\rm sech}}
\newcommand{\half}{{\tfrac{1}{2}}}

\newcommand{\nlb}{\protect\nolinebreak}

\newcommand{\Myr}{\nlb\mbox{\,Myr}}

\newcommand{\km}{\nlb\mbox{\,km}}
\newcommand{\cm}{\nlb\mbox{\,cm}}
\newcommand{\cmmm}{\nlb\mbox{\,cm$^{-3}$}}

\newcommand{\sm}{\nlb\mbox{\,s$^{-1}$}}

\newcommand{\erg}{\nlb\mbox{\,erg}}
\newcommand{\gauss}{\nlb\mbox{\,G}}
\newcommand{\Kelvin}{\nlb\mbox{\,K}}

\newcommand{\Bf}{{mag\-netic field}}
\newcommand{\Bfs}{{mag\-netic fields}}
\newcommand{\Dsi}{{Density-shear in\-stability}}
\newcommand{\dsi}{{density-shear in\-stability}}

\begin{document}

\title{\Dsi\ in electron MHD}%
\author{T.~S.~Wood}%
\email{t.wood@leeds.ac.uk}%
\author{R.~Hollerbach}%
\affiliation{Department of Applied Mathematics, University of Leeds, Leeds, LS2 9JT, United Kingdom}%
\author{M.~Lyutikov}%
\affiliation{Department of Physics, Purdue University, 525 Northwestern Avenue, West Lafayette, Indiana 47907-2036, USA}%

\date{\today}%

\begin{abstract}
We discuss a novel instability in inertia-less electron magneto-hydrodynamics (EMHD),
which arises from a combination of electron velocity shear and electron density gradients.
The unstable modes have a
lengthscale longer than the transverse density scale,
and a growth-rate of the order of the inverse Hall timescale.
We suggest that this \dsi\ may be of importance
in magnetic reconnection regions on scales smaller than the ion skin depth, and
in neutron star crusts.
We demonstrate that the so-called Hall drift instability,
previously argued to be relevant in neutron star crusts,
is a resistive tearing instability rather than an instability of the Hall term itself.
We argue that the \dsi\ is of greater significance in neutron stars than the tearing instability,
because it generally has a faster growth-rate and is less sensitive to geometry and boundary conditions.
We prove that, for uniform electron density, EMHD is ``at least as stable'' as regular, incompressible MHD,
in the sense that any field configuration that is stable in MHD is also stable in EMHD.
We present a connection between the \dsi\ in EMHD
and the magneto-buoyancy instability in anelastic MHD.
\end{abstract}



\maketitle%

\section{Introduction}

Electron magneto-hydrodynamics (EMHD)
is a regime of plasma physics in which positive and neutral particles are approximately immobile,
so that only the dynamics of the (lighter) electrons needs to be considered
\cite{Kingsep-etal87,Gordeev-etal94}.
The flow of electrons induces \Bfs\ via the Hall effect,
and the \Bf\ influences the electrons via the Lorentz force.
EMHD was first studied in the context of laboratory plasma experiments \cite{GordeevRudakov69},
where it is applicable on scales smaller than the ion skin depth,
and has also been used to explain (nearly) collisionless reconnection in the solar corona and magnetotail
\cite{Mandt-etal94,Avinash-etal98,DengMatsumoto01}.

Another important application of EMHD is in neutron stars \cite{GoldreichReisenegger92,Cumming-etal04}.
Within the outermost 1km of a neutron star, called the crust,
the ions are locked into a solid lattice, 
and the dynamics of the \Bf\ is therefore governed by the EMHD equations.
The structure and topology of the \Bf\ determines the pattern of radiation from the star,
and thus its observational signature, thermal evolution, and spin-down timescale \cite{Pacini67,*Gold69}.
Furthermore, the evolution of the \Bf\ in the crust can trigger radiation bursts and flares,
either via internal crustal failure \cite{ThompsonDuncan95,*ThompsonDuncan96,LevinLyutikov12}
or by twisting the external \Bf\ lines until they undergo fast reconnection \cite{Lyutikov03,Lyutikov06}.

An important question then is whether \Bfs\ in EMHD have a preferred structure,
and whether some field configurations are unstable.
Although there have been many studies of instability and turbulence in EMHD,
almost all of the known instabilities require either finite ohmic resistivity \cite{Gordeev70}
or finite electron inertia \cite{Bulanov-etal92,Califano-etal99,GaurDas12}.
In the crust of a neutron star, however,
electron inertia is entirely negligible in comparison with the Lorentz and Coulomb forces.
Furthermore, most studies of EMHD turbulence have only considered the structure of the field in spectral space
\cite{Biskamp-etal96,Biskamp-etal99,Dastgeer-etal00,ChoLazarian04,Galtier08b}.
More recently, numerical simulations have suggested that the \Bf\ in a neutron star crust
evolves towards a quasi-equilibrium, ``frozen-in'' state on a relatively short timescale ($\lesssim 1\Myr$)
\cite{HollerbachRudiger02,HollerbachRudiger04,PonsGeppert07,WareingHollerbach09a,WareingHollerbach10,GC14}.
However, these studies are all either two-dimensional, or else neglect variations in the electron density.
Whether the quasi-equilibrium states found in these studies would be dynamically stable under more realistic conditions is unknown.

\defcitealias{RheinhardtGeppert02}{RG02}%
\citet{RheinhardtGeppert02},
\citetalias{RheinhardtGeppert02} hereafter,
have presented numerical results demonstrating an instability in the inertia-less EMHD equations,
which they argue is caused by Hall drift in the presence of a non-uniform \Bf,
and which they called the ``Hall drift instability''. %
However, they find that the growth-rate of the instability is very sensitive to the choice of boundary conditions,
and also seems to vanish in the limit of high electrical conductivity.
These observations suggest that the instability may, in fact,
be a resistive tearing mode rather than an instability of the Hall term itself.
The distinction is important, because tearing instability occurs only for rather particular \Bf\ configurations,
which calls into question the general applicability of
\citetalias{RheinhardtGeppert02}'s results to neutron stars.
Tearing instability generally occurs only if the \Bf\ has a null surface,
within which the ideal EMHD equations become singular.

The goal of this paper is to determine the stability properties of inertia-less EHMD equilibrium states,
for both finite and infinite electrical conductivity.
We also consider the effect of non-uniform electron density,
which was neglected in \citetalias{RheinhardtGeppert02}'s
original model, but included in a subsequent work \cite{Rheinhardt-etal04}. %
This effect is almost certainly important in real neutron star crusts,
in which the electron density scale-height
is typically only a few percent of the crust thickness \cite{ChamelHaensel08}.
We show that electron velocity shear together with density gradients can produce an instability,
which resembles an instability described originally in Ref.~\onlinecite{GordeevRudakov69}.
We suggest that this \dsi\ was present in the results of Ref.~\onlinecite{Rheinhardt-etal04}, %
which explains the significant discrepancies between their results and those %
of \citetalias{RheinhardtGeppert02}. %
We present an explicit, analytical instance of the \dsi, and discuss its implications for the
evolution of \Bfs\ in neutron star crusts.

\section{The inertia-less EMHD equations}
The basic equations describing EMHD
are
\begin{align}
  \frac{\partial\mathbf{B}}{\partial t} &= -c\nablab\times\mathbf{E} \label{eq:Faraday} \\
  \mathbf{J} &= \frac{c}{4\pi}\nablab\times\mathbf{B} \label{eq:Ampere} \\
  \frac{\ee^2n}{\sigma}\mathbf{v}
  &=
  -\frac{1}{n}\nablab P - \ee\left(\mathbf{E} + \frac{\mathbf{v}}{c}\times\mathbf{B}\right). \label{eq:balance}
\end{align}
These are, respectively, Faraday's law, Amp\`ere's law, and the force balance for the electron fluid
in Gaussian cgs units.
Here,
$\mathbf{E}$ and $\mathbf{B}$ are the electric and \Bfs, $\mathbf{J}$ is the electric current,
$\mathbf{v}$ is the electron fluid velocity, $n$ is the electron number density,
$P$ is the electron pressure, $\sigma$ is the electrical conductivity,
$c$ is the speed of light, and $\ee = |\ee|$ is the elementary charge.
Note that we neglect the electron inertia in \eq(\ref{eq:balance}), which is negligible on scales much larger
than the electron skin depth, $d = (c/\ee)(m^\star/4\pi n)^{1/2}$, where $m^\star$ is the effective mass of the
electrons.  In the crust of a neutron star the skin depth is tiny, $d \sim 10^{-11}\cm$,
making this an excellent approximation \cite{ChamelHaensel08}.
As noted earlier,
the absence of electron inertia has important implications for
EMHD stability, because most known instabilities in EMHD arise from inertial effects.

The electric current produced by the flow of electrons is
\begin{equation}
  \mathbf{J} = -\ee n\mathbf{v},
  \label{eq:J}
\end{equation}
and substituting this relation into \eq(\ref{eq:balance}) yields a generalized Ohm's law.
\Eqs(\ref{eq:Ampere}) and (\ref{eq:J}) together imply that the electron density $n$ is steady in the Eulerian description, because
\begin{align}
  \frac{\partial n}{\partial t} &= -\nablab\cdot(n\mathbf{v}) \nonumber \\
  &= \frac{c}{4\pi\ee}\nablab\cdot(\nablab\times\mathbf{B}) \nonumber \\
  &= 0.
\end{align}
\Eqs(\ref{eq:Faraday})--(\ref{eq:J}) can be combined into a single equation for the evolution of the \Bf,
\begin{equation}
  \frac{\partial\mathbf{B}}{\partial t}
  =
  \nablab\times\left[
    \frac{c}{4\pi\ee n}\mathbf{B}\times(\nablab\times\mathbf{B}) - \frac{c^2}{4\pi\sigma}\nablab\times\mathbf{B}
  \right]
  + \frac{c}{\ee n^2}\nablab P\times\nablab n.
  \label{eq:dimensional}
\end{equation}
The three terms on the right-hand side are, respectively, the Hall effect, ohmic decay, and the Biermann battery effect.
From here on,
we follow Ref.~\onlinecite{GoldreichReisenegger92} and \citetalias{RheinhardtGeppert02} in neglecting the Biermann battery term,
which represents the generation of electric currents by electron baroclinicity.
This term becomes negligible at sufficiently low temperatures, for which the electron gas is fully degenerate
and therefore barotropic.  More precisely, the Biermann term is negligible in comparison with the Hall term if
\begin{equation}
  (T/T_{\rm F})^2 \ll \frac{|\mathbf{B}|^2}{4\pi\epsilon_{\rm F}n}, \label{eq:no_battery}
\end{equation}
where $\epsilon_{\rm F}$ is the Fermi energy and $T_{\rm F} = \epsilon_{\rm F}/k_{\rm B}$ is the Fermi temperature.
Typical orders of magnitude for the various parameters in a neutron star crust are
\cite{Cumming-etal04}
$n \sim 10^{35}\cmmm$,
$\epsilon_{\rm F} \sim 10^{-4}\erg$,
$|\mathbf{B}| \sim 10^{13}\gauss$,
and $\sigma \sim 10^{25}\sm$.
Condition~(\ref{eq:no_battery}) is then $T \ll 10^9\Kelvin$,
which is satisfied in isolated neutron stars older than about 1\Myr.

The relative importance of the Hall and ohmic decay terms in \eq(\ref{eq:dimensional}) is measured by the Hall parameter,
\begin{equation}
  H = \frac{\sigma}{c\ee n}|\mathbf{B}|.
\end{equation}
For the parameter values just given we find $H \sim 10^2$, implying that the Hall term dominates
the evolution of the \Bf.
Assuming a typical lengthscale $L \sim 1\km$,
comparable to the thickness of the crust,
the characteristic timescale for \Bf\ evolution is then
\begin{equation}
  t_{\rm Hall} = \frac{4\pi\ee nL^2}{c|\mathbf{B}|} \sim 1\Myr.
\end{equation}

In what follows, we work exclusively with non-dimensional quantities.
We use $L$ and $t_{\rm Hall}$ as units of length and time respectively,
and we measure $\mathbf{B}$ and $n$ in units of the characteristic values given above.
In these units, \eq(\ref{eq:dimensional}) becomes
\begin{equation}
  \frac{\partial\mathbf{B}}{\partial t}
  =
  \nablab\times\left[
    \frac{1}{n}\mathbf{B}\times\mathbf{J} - \eta\mathbf{J}
  \right],
  \label{eq:EMHD}
\end{equation}
where $\eta$ is a dimensionless diffusivity of order $H^{-1} \ll 1$,
and where
\begin{equation}
  \mathbf{J} = \nablab\times\mathbf{B}
\end{equation}
is the dimensionless electric current.
We will consider linear perturbations to equilibrium states,
\ie\ to steady solutions of \eq(\ref{eq:EMHD}).
The perturbation to the \Bf, $\delta\mathbf{B}$,
obeys the linear equation
\begin{equation}
  \frac{\partial}{\partial t}\delta\mathbf{B}
  =
  \nablab\times\left[
    - \frac{1}{n}\mathbf{J}\times\delta\mathbf{B}
    + \frac{1}{n}\mathbf{B}\times(\nablab\times\delta\mathbf{B})
    - \eta\nablab\times\delta\mathbf{B}
  \right].
  \label{eq:linear}
\end{equation}
The first term on the right-hand side of \eq(\ref{eq:linear}) represents advection of the perturbations by the
electron velocity,
the second term gives rise to whistler waves (a.k.a.~helicons) for a uniform background field and uniform density
\cite{BoswellChen97,Newell10},
and the third term represents resistive diffusion of the perturbation.

From \eq(\ref{eq:EMHD}) it can be shown that
\begin{equation}
  \frac{\dd}{\dd t}\int_V\!\dd V\,\tfrac{1}{2}|\mathbf{B}|^2
  =
  - \int_V\!\dd V\,\eta|\mathbf{J}|^2
  - \int_{\partial V}\!\dd S\,\hat{\mathbf{n}}\cdot\mathbf{B}\times\left[
    \frac{1}{n}\mathbf{B}\times\mathbf{J} - \eta\mathbf{J}
  \right]
  \label{eq:energy}
\end{equation}
for any volume $V$ with boundary $\partial V$ and outward normal $\hat{\mathbf{n}}$.
For a closed system the surface integral vanishes, and so the magnetic energy decays in time,
at a rate that depends on the diffusivity $\eta$.
This does not, however, imply that the system is stable, because perturbations can still grow
by extracting energy from the background \Bf.
In fact, it can be shown from \eq(\ref{eq:linear}) that, omitting surface integrals,
\begin{align}
  \frac{\dd}{\dd t}\int_V\!\dd V\,\tfrac{1}{2}|\delta\mathbf{B}|^2
  &=
  \int_V\!\dd V\,\left[
    \delta B_i\left(\nablab\cdot(\mathbf{J}/n)\frac{\delta_{ij}}{2} - \frac{\partial(J_i/n)}{\partial x_j}\right)\delta B_j
    - \eta|\delta\mathbf{J}|^2
  \right].
  \label{eq:linear_energy}
\end{align}
So, in a closed system, the only possible source of instability is spatial gradients in $\mathbf{J}/n$,
\ie\ gradients in the electron velocity.

\section{Uniform electron density}
\label{sec:uniform}

We begin by considering the simplest case of EMHD,
in which $\eta=0$ and $n$ is constant.  Without loss of generality we take $n=1$
--- any other constant value can be obtained simply by rescaling the \Bf.
We consider linear perturbations $\delta\mathbf{B}$ to a steady background field $\mathbf{B}$,
which is a solution of the equation
\begin{equation}
  0 = \nablab\times\left[\mathbf{B}\times(\nablab\times\mathbf{B})\right].
  \label{eq:ideal_equilibrium}
\end{equation}
The linear equation for the perturbations (\ref{eq:linear}) simplifies in this case to
\begin{equation}
  \frac{\partial}{\partial t}\delta\mathbf{B}
  =
  \nablab\times\left[
    - \mathbf{J}\times\delta\mathbf{B}
    + \mathbf{B}\times(\nablab\times\delta\mathbf{B})
  \right],
  \label{eq:linear_ideal}
\end{equation}
and \eq(\ref{eq:linear_energy}) becomes
\begin{equation}
  \frac{\dd}{\dd t}\int_V\!\dd V\,\tfrac{1}{2}|\delta\mathbf{B}|^2
  =
  - \int_V\!\dd V\,\delta B_i\frac{\partial J_i}{\partial x_j}\delta B_j.
  \label{eq:shear}
\end{equation}
So a necessary condition for instability is the presence of electron velocity shear.

\subsection{Stability of straight field lines}
\label{sec:straight}

Following \citetalias{RheinhardtGeppert02}, we now
adopt a Cartesian coordinate system $(x,y,z)$ and
consider a background \Bf\ of the form
\begin{equation}
  \mathbf{B} = B(z)\,\mathbf{e}_x
  \label{eq:1d}
\end{equation}
where $\mathbf{e}_x$ is the unit vector in the $x$ direction.
A field of this form satisfies the equilibrium condition (\ref{eq:ideal_equilibrium}) for any choice of the function $B(z)$.
Since the background field depends only on $z$, we may seek eigenmode solutions to the perturbation equation (\ref{eq:linear_ideal}) of the form
\begin{equation}
  \delta\mathbf{B} = \mathbf{b}(z)\exp(\lambda t + \ii k_xx + \ii k_y y)
  \label{eq:ansatz}
\end{equation}
where $\lambda$ is the (possibly complex) growth-rate and $\mathbf{b}$ is a complex amplitude function.
Substituting (\ref{eq:1d}) and (\ref{eq:ansatz}) into \eq(\ref{eq:linear_ideal}),
we eventually obtain a single equation for the $z$ component of $\mathbf{b}$:
\begin{equation}
  \frac{b_z''}{b_z} = \frac{B''}{B} + k_x^2 + k_y^2
    + \left(\frac{\lambda}{Bk_x} - \ii\frac{B'k_y}{Bk_x}\right)\frac{\lambda}{Bk_x},
  \label{eq:2d_linear}
\end{equation}
where primes denote derivatives with respect to $z$.
For the moment, we restrict attention to perturbations that have $k_y=0$.
This equation then becomes simply
\begin{equation}
  \frac{b_z''}{b_z} = \frac{B''}{B} + k_x^2
    + \frac{\lambda^2}{B^2k_x^2}.
  \label{eq:1d_linear}
\end{equation}
By analogy with the one-dimensional Schr\"odinger equation for a potential well,
we see that in order for this equation to have bounded solutions
the right-hand side must be negative for some range of $z$.
Therefore a necessary condition for instability (\ie\ ${\rm Re}\{\lambda\} > 0$)
is that the term $B''/B$ must be negative for some range of $z$.
\citetalias{RheinhardtGeppert02} argue that there will be unstable modes provided that
$B''/B$ is chosen to be sufficiently negative (see also Ref.~\onlinecite{Cumming-etal04}, \S4.3.5).
However, this reasoning is flawed, as can be seen by multiplying \eq(\ref{eq:1d_linear}) by $|b_z|^2$
and integrating in $z$.  We then obtain
\begin{equation}
  \left[b_z^\star\left(b_z' - \frac{B'}{B}b_z\right)\right] = \int\!\dd z\,\left|b_z' - \frac{B'}{B}b_z\right|^2
  + \int\!\dd z\,\left(k_x^2 + \frac{\lambda^2}{B^2k_x^2}\right)|b_z|^2,
  \label{eq:parts}
\end{equation}
where $b_z^\star$ denotes the complex conjugate of $b_z$.
If the boundary conditions are chosen so that the left-hand side of \eq(\ref{eq:parts}) vanishes, then we deduce that
\begin{equation}
  - \lambda^2 = \dfrac{\displaystyle k_x^2\int\!\dd z\,\left|b_z' - \dfrac{B'}{B}b_z\right|^2
  + k_x^4\int\!\dd z\,|b_z|^2}{\displaystyle\int\!\dd z\,\dfrac{|b_z|^2}{B^2}},
  \label{eq:bound}
\end{equation}
implying that the growth-rate $\lambda$ is purely imaginary, and so all modes are neutrally stable.
This conclusion also applies for practically any other sensible choice of boundary conditions.
For example, suppose that our domain is $-1<z<1$, and that the region outside the domain is a vacuum.
The background field must then have $B'(z)=0$ at $z=\pm1$, and the boundary conditions for the
perturbations are $b_z' = \mp|k_x|b_z$ at $z=\pm1$.  The left-hand side of \eq(\ref{eq:parts})
is then negative, and the conclusion that $\lambda$ is imaginary holds even more strongly.
Allowing for $k_y\neq0$ does not alter this conclusion; in fact, it can be shown from \eq(\ref{eq:2d_linear})
that the generalization of \eq(\ref{eq:bound}) when $k_y\neq0$ is
\begin{equation}
  - \left[\lambda - \tfrac{1}{2}\ii k_y\dfrac{\displaystyle\int\!\dd z\,\frac{B'}{B^2}|b_z|^2}{\displaystyle\int\!\dd z\,\frac{|b_z|^2}{B^2}}\right]^2 =
  \dfrac{\displaystyle k_x^2\int\!\dd z\,\left|b_z' - \dfrac{B'}{B}b_z\right|^2
  + k_x^2(k_x^2+k_y^2)\int\!\dd z\,|b_z|^2}{\displaystyle\int\!\dd z\,\dfrac{|b_z|^2}{B^2}}
  + \left[\tfrac{1}{2}k_y\dfrac{\displaystyle\int\!\dd z\,\frac{B'}{B^2}|b_z|^2}{\displaystyle\int\!\dd z\,\frac{|b_z|^2}{B^2}}\right]^2.
  \label{eq:bound2}
\end{equation}
So $k_y$ produces a frequency splitting between modes that propagate ``upstream'' and ``downstream'' with respect
to the electron velocity, but all modes remain purely oscillatory, and therefore the system is stable.

It is straightforward to generalize these results still further by considering a \Bf\ of the form
$\mathbf{B} = B_x(z)\,\mathbf{e}_x + B_y(z)\,\mathbf{e}_y$,
which automatically satisfies the equilibrium condition (\ref{eq:ideal_equilibrium}).
\Eq(\ref{eq:2d_linear}) then becomes
\begin{equation}
  \frac{b_z''}{b_z} = \frac{\mathbf{B}''\cdot\mathbf{k}}{\mathbf{B}\cdot\mathbf{k}} + |\mathbf{k}|^2
    + \left(\frac{\lambda}{\mathbf{B}\cdot\mathbf{k}} - \ii\frac{[\mathbf{B}'\times\mathbf{k}]_z}{\mathbf{B}\cdot\mathbf{k}}\right)
      \frac{\lambda}{\mathbf{B}\cdot\mathbf{k}},
  \label{eq:3d_linear}
\end{equation}
from which stability can be demonstrated as before.
In this way, it can be shown that any field with straight field lines (\ie\ with $\mathbf{B}\cdot\nablab\mathbf{B} = 0$)
is stable in EMHD if the electron density is uniform.
This result can also be obtained by a more mathematically elegant argument, as shown later in
\S\ref{sec:Lundquist}.

We note that this result contradicts Ref.~\onlinecite{Drake-etal94},
which claimed to have demonstrated instability for a \Bf\ with straight field lines.
Although their derivation assumes finite electron inertia, their instability remains even in the limit of zero inertia.
We suggest that there are three reasons why they obtained this incorrect result.
First, they assumed a background \Bf\ that varies in one direction, but they only considered perturbations
that are invariant in that direction.  This is equivalent to seeking solutions of \eq(\ref{eq:3d_linear})
for which the left-hand side vanishes, which is obviously impossible for any non-trivial \Bf.
Second, they prescribed a uniform background \Bf\ and a non-uniform background current,
which is incompatible with Amp\`ere's law (\ref{eq:Ampere}).
Third, they considered a localized region in which the current changes sign, and therefore neglected terms involving
the current while retaining terms involving its spatial gradient.  There is no rigorous basis for this approximation.
We believe that these inconsistencies explain the contradiction between their results and ours.
We emphasize, however, that there \emph{are} instabilities in the EMHD equations when the fluid has finite inertia.
This can be proved by simply observing that, on scales much smaller than the electron skin depth,
the EMHD equations become identical to the equations for an incompressible, non-magnetic fluid
\cite{Califano-etal99,GaurDas12},
for which there are numerous well-known instabilities.

\subsection{Resistive tearing instability}
\label{sec:tearing}
The argument presented in the last section assumes, implicitly,
that all of the integrals in \eqs(\ref{eq:parts})--(\ref{eq:bound2}) are well defined.
However, if the background field in \eq(\ref{eq:1d}) is chosen such that $B(z)$ vanishes somewhere within the domain
then these integrals may be singular.  We see from \eqs(\ref{eq:2d_linear}) and (\ref{eq:1d_linear})
that the eigenmodes are also singular in such cases.
This suggests that reintroducing finite resistivity, $\eta > 0$, %
will significantly alter the structure of the eigenmodes,
as well as their stability properties.
In particular, by analogy with ``regular'' MHD,
we anticipate that such fields can be subject to resistive tearing instabilities \cite{Furth-etal63,PegoranoSchep86}.
The existence of tearing instabilities in EMHD has been convincingly demonstrated previously
\cite{Gordeev70,Bulanov-etal92,FruchtmanStrauss93},
so here we simply summarize the essential points of the analysis,
and compare the predictions of the theory with the results of \citetalias{RheinhardtGeppert02}.
For simplicity, we also restrict attention to modes of the form given by \eq(\ref{eq:ansatz}) with $k_y=0$;
the fastest growing mode found by \citetalias{RheinhardtGeppert02} was in this category.

Suppose that $B(z)$ vanishes for some value of $z$, say $z=z_0$.
If resistive diffusion is sufficiently weak then we expect \eq(\ref{eq:1d_linear}) to hold to a good approximation away
from the singularity at $z_0$.
We refer to the solution of this equation as the ``outer'' solution.
In a neighborhood of the singularity we
approximate $B(z) = B'(z_0)(z-z_0)$,
and we assume that %
$\eta$ is constant.
Substituting the ansatz (\ref{eq:ansatz}) into \eq(\ref{eq:linear}),
we thus obtain a pair of coupled equations for $b_y$ and $b_z$,
\begin{align}
  (\lambda-\eta\nabla^2) b_y &= B'(z-z_0)\nabla^2 b_z \label{eq:inner1} \\
  (\lambda-\eta\nabla^2) b_z &= B'(z-z_0)k_x^2 b_y, \label{eq:inner2}
\end{align}
whose solution we will call the ``inner'' solution.
For now, we assume that $B'$ and $k_x$ are of order unity, and that $\eta$ is of order $H^{-1} \ll 1$.
We then find that resistive diffusion %
only becomes important on scales smaller than $\sqrt{\eta/|B'k_x|} \sim H^{-1/2}$,
and we therefore approximate $\nabla^2 \simeq \partial^2/\partial z^2$ in these equations.
We seek solutions for which $ b_y$ is antisymmetric about $z_0$ and $ b_z$ is symmetric.
Following Ref.~\onlinecite{Furth-etal63} we anticipate that, provided $|\lambda| \ll |B'k_x|$,
both $ b_y$ and $ b_z$ will be approximately constant in a neighborhood of $z_0$.
This allows us to approximate\footnote{%
In the ``regular'' MHD case considered in Ref.~\onlinecite{Furth-etal63} this approximation can be rigorously justified
by asymptotic analysis.  Whether there is such a rigorous justification in the EMHD case is unclear,
but a more careful study in Ref.~\onlinecite{Attico-etal00} suggests that this approximation is reasonable.}
\eqs(\ref{eq:inner1}) and (\ref{eq:inner2}) as
\begin{align}
  -\eta\frac{\partial^2}{\partial z^2} b_y &= B'(z-z_0)\frac{\partial^2}{\partial z^2} b_z
  \label{eq:inner3} \\
  \lambda b_{z0} - \eta\frac{\partial^2}{\partial z^2} b_z &= B'(z-z_0)k_x^2 b_y,
  \label{eq:inner4}
\end{align}
where $ b_{z0}$ is the value of $ b_z$ at $z=z_0$.
Finally, we introduce a stretched coordinate $\tilde{z} = \left|\dfrac{2B'k_x}{\eta}\right|^{1/2}(z-z_0)$,
which leads to the following equation for the quantity
$b(\tilde{z}) = \dfrac{B'}{\lambda b_{z0}}\left|\dfrac{2k_x^3\eta}{B'}\right|^{1/2} b_y$:
\begin{align}
  b'' - \tfrac{1}{4}\tilde{z}^2b + \tfrac{1}{2}\tilde{z} &= 0.
  \label{eq:Furth}
\end{align}
This is a special case of the equation obtained in Ref.~\onlinecite{Furth-etal63},
and so we can use their results from here on.
The unique regular solution of \eq(\ref{eq:Furth}) can be expressed as a sum of Hermite functions,
\begin{equation}
  b(\tilde{z}) = \sum_{\mbox{odd $n$}}\frac{2}{n+\tfrac{1}{2}}
    \left[\frac{\Gamma(\tfrac{1}{2}n+1)}{\Gamma(\tfrac{1}{2}n+\tfrac{1}{2})}\right]^{1/2}
    \frac{(-1)^n\ee^{\tilde{z}^2\!/4}}{\pi^{1/4}(n!)^{1/2}}\frac{\dd^n}{\dd\tilde{z}^n}\ee^{-\tilde{z}^2\!/2}.
  \label{eq:regular}
\end{equation}
Although $ b_z$ is roughly constant in the inner solution, its $z$ derivative changes rapidly across $z=z_0$,
by an amount
\begin{align}
  \Delta' &\equiv \frac{1}{ b_{z0}}\int\!\dd z\,\frac{\partial^2}{\partial z^2} b_z \nonumber \\
  &= \frac{\lambda}{\sqrt{|2B'k_x\eta|}}\int_{-\infty}^{+\infty}\dd\tilde{z}\,(1-\tfrac{1}{2}\tilde{z}b(\tilde{z})) \nonumber \\
  &= 2\pi\frac{\Gamma(3/4)}{\Gamma(1/4)}\frac{\lambda}{\sqrt{|B'k_x\eta|}}.
\end{align}
Instability therefore requires that $\Delta' > 0$.
The approximations made in \eqs(\ref{eq:inner3}) and (\ref{eq:inner4}) are self-consistent provided that
$1/|\Delta'|$ is much larger than the width of the inner region, \ie\
\begin{align*}
  \left|\frac{\eta}{B'k_x}\right|^{1/2} &\ll 1/|\Delta'| \\
  \Leftrightarrow |\lambda| &\ll |B'k_x|
\end{align*}
as expected.

\citetalias{RheinhardtGeppert02}
take a background field of the form $B(z) = B_0(1-z^2)$,
with $B_0$ a constant,
over the domain $-1<z<1$,
and use ``vacuum'' boundary conditions for the perturbations,
\ie\ $b_y=0$ and $\dfrac{\partial}{\partial z} b_z = \mp|k_x| b_z$ at $z=\pm1$.
Because the singularities in their case occur exactly at the boundaries of the domain,
the change in the derivative of $b_z$ across the inner regions is $\tfrac{1}{2}\Delta'$.
The effective boundary conditions for the outer solution in their case are therefore
\begin{equation}
  \dfrac{\partial}{\partial z} b_z = \mp(|k_x| + \tfrac{1}{2}\Delta') b_z \;\;\; \mbox{at} \;\;\; z=\pm1.
  \label{eq:outer_BC}
\end{equation}
We are now in a position to calculate the dispersion relation for the EMHD tearing modes.
Rather than taking \citetalias{RheinhardtGeppert02}'s field $B(z) = B_0(1-z^2)$,
we use a very similar field $B(z) = B_0\cos(\tfrac{\pi}{2}z)$
that allows us to calculate the outer solution analytically.
The general solution of \eq(\ref{eq:1d_linear}) can then be written in terms of associated Legendre functions,
\begin{flalign}
  && b_z &= \cos^{1/2}(\tfrac{\pi}{2}z)P_l^m(\sin(\tfrac{\pi}{2}z)),& \\
  & \mbox{where} & m^2 &= \frac{1}{4} + \left(\frac{2\lambda}{\pi B_0k_x}\right)^2&
  \label{eq:cond_m} \\
  & \mbox{and} & \left(l+\frac{1}{2}\right)^2 &= 1 - \left(\frac{2k_x}{\pi}\right)^2.&
  \label{eq:cond_l}
\end{flalign}
The only outer solutions that are bounded for all $z$ are those with $m\leqslant\tfrac{1}{2}$,
implying that the growth-rate $\lambda$ would be imaginary in the absence of singularities.
We now seek tearing solutions, which are marginally stable
(\ie\ $\lambda=0$) outer solutions that can be matched to the boundary conditions (\ref{eq:outer_BC})
with $\Delta' > 0$.
The marginally stable outer solutions form odd and even families,
$b_z = \sin(\tfrac{\pi^2}{4}-k_x^2)^{1/2}z$
and $b_z = \cos(\tfrac{\pi^2}{4}-k_x^2)^{1/2}z$
respectively.
Applying the boundary conditions (\ref{eq:outer_BC})
leads to distinct dispersion relations for the two families:
\begin{align}
  \mbox{odd:} \;\;\;
  \frac{\Gamma(3/4)}{\Gamma(1/4)}\left(\frac{2\pi}{\eta B_0|k_x|}\right)^\half\lambda
  &= -\left(\tfrac{\pi^2}{4}-k_x^2\right)^\half\cot\left(\tfrac{\pi^2}{4}-k_x^2\right)^\half - |k_x|
  \label{eq:odd} \\
  \mbox{even:} \;\;\;
  \frac{\Gamma(3/4)}{\Gamma(1/4)}\left(\frac{2\pi}{\eta B_0|k_x|}\right)^\half\lambda
  &= \left(\tfrac{\pi^2}{4}-k_x^2\right)^\half\tan\left(\tfrac{\pi^2}{4}-k_x^2\right)^\half - |k_x|.
  \label{eq:even}
\end{align}
The odd modes are stable for all $k_x$, but the even modes are unstable for $k_x\lesssim1$.
Away from the boundaries, the structure of the unstable modes, given by the outer solution, is
\begin{equation}
  \delta\mathbf{B} \propto \left(\begin{array}{c}
    (\tfrac{\pi^2}{4}-k_x^2)^{1/2} \\
    0 \\
    \ii k_x\cot\left[(\tfrac{\pi^2}{4}-k_x^2)^{1/2}z\right]
  \end{array}\right)\sin\left[(\tfrac{\pi^2}{4}-k_x^2)^{1/2}z\right]\exp(\ii k_xx + \lambda t)
\end{equation}
so the instability is purely two-dimensional, except near the boundaries, and undular in $x$.
The growth-rate given by \eq(\ref{eq:even}) diverges as $|k_x|\to0$, with $\lambda \sim (\eta B_0)^{1/2}|k_x|^{-3/2}$,
but the self-consistency condition $\lambda \ll B_0|k_x|$ is
evidently violated in this limit.
We expect the actual fastest growing mode to have $\lambda \sim B_0|k_x| \sim (\eta B_0)^{1/2}|k_x|^{-3/2}$ in the limit $(\eta/B_0) \sim H^{-1} \to 0$,
and so $\lambda \sim \eta^{1/5}B_0^{4/5}$ and $|k_x| \sim (\eta/B_0)^{1/5}$.

Although we are not able to analytically predict the exact wavenumber and growth-rate for the background field
studied by \citetalias{RheinhardtGeppert02}, we can make the following predictions regarding
the most unstable tearing mode in their case:
\begin{enumerate}
  \item in the bulk of the domain, the instability is undular and two-dimensional,
  with a wavenumber $\sim (\eta/B_0)^{1/5}$ in the field-wise direction;
  \item the eigenmode adjusts rapidly to meet the boundary conditions, within a layer of width $\sim (\eta/B_0)^{2/5}$;
  \item the growth-rate $\lambda$
  vanishes as $\eta \to 0$, roughly as $\eta^{1/5}B_0^{4/5}$;
  \item the growth-rate is very sensitive to boundary conditions.
\end{enumerate}
The instability found numerically by \citetalias{RheinhardtGeppert02} agrees with each of these predictions.
In particular, they find that $\delta B_y \simeq 0$, except in thin boundary layers, and that the growth-rate
vanishes in the limit $\eta \to 0$ with $\lambda \sim \eta(B_0/\eta)^q$ and $q\in(0.7,0.9)$.
These are strong indications that the instability is driven by diffusion across the
singularities where $|\mathbf{B}| = 0$, rather than by the Hall effect itself.

In the light of the above, we might question whether the instability found by \citetalias{RheinhardtGeppert02}
has much relevance to neutron stars.  The tearing instability requires rather specific conditions,
and it is not clear that these would arise naturally.  In fact, the particular choice of background \Bf\
$\mathbf{B} = B_0(1-z^2)\mathbf{e}_x$
used by \citetalias{RheinhardtGeppert02} is rather artificial,
and does not have an obvious analogue in spherical geometry.
In their model, $z$ is intended to be the vertical coordinate,
and therefore becomes the radial coordinate in spherical geometry.
Because there are no spherical EMHD equilibrium states with a purely toroidal field
\cite{Gourgouliatos-etal13},
we must interpret $x$ as the latitudinal coordinate.
But in a sphere with vacuum outer boundary conditions, the latitudinal component of the field would
not be expected to vanish at the outer boundary, and so there would be no tearing instability there.
Nor is the latitudinal component expected to vanish at the bottom of the crust,
at the boundary with the superconducting outer core.

\subsection{Connection with incompressible MHD}
\label{sec:Lundquist}
If the instability found by \citetalias{RheinhardtGeppert02} is in fact a resistive tearing instability,
then there remains the question of whether there exist any instabilities in ideal (\ie\ non-resistive) inertia-less EMHD.
In analyzing this question, it proves useful to use a new linear perturbation variable in place of $\delta\mathbf{B}$.
We therefore introduce the Lagrangian perturbation $\bm{\xi}$, which we define as the difference between
the Eulerian position of a Lagrangian particle in the perturbed and unperturbed systems.
The (Eulerian) perturbations to the \Bf\ and electron density can then be expressed as
\begin{align}
  \delta\mathbf{B} &= \nablab\times(\bm{\xi}\times\mathbf{B})
  \label{eq:deltaB} \\
  \delta n &= - \nablab\cdot(n\bm{\xi}).
\end{align}
The Lagrangian perturbation is kinematically related to the electron velocity as
\begin{align}
  \frac{\partial}{\partial t}(n\bm{\xi}) &= \delta(n\mathbf{v}) + \nablab\times(n\mathbf{v}\times\bm{\xi}).
\end{align}
We close the equations using the (dimensionless) relations
\begin{align}
  \mathbf{J} = \nablab\times\mathbf{B} \;\;\; &\Rightarrow \;\;\; \delta\mathbf{J} = \nablab\times\delta\mathbf{B} \\
  \mathbf{J} = - n\mathbf{v} \;\;\; &\Rightarrow \;\;\; \delta\mathbf{J} = - \delta(n\mathbf{v})
\end{align}
which imply that
\begin{align}
  n\frac{\partial\bm{\xi}}{\partial t} &= \nablab\times\left[\nablab\times(\mathbf{B}\times\bm{\xi}) - \mathbf{J}\times\bm{\xi}\right]
  \label{eq:xi} \\
  \nablab\cdot(n\bm{\xi}) &= 0.
  \label{eq:div_nxi}
\end{align}
Taking the background density to be $n=1$, the linear equations for $\bm{\xi}$ become
\begin{align}
  \frac{\partial\bm{\xi}}{\partial t} &= \nablab\times\left[\nablab\times(\mathbf{B}\times\bm{\xi}) - \mathbf{J}\times\bm{\xi}\right] 
  \label{eq:ideal_xi} \\
  \nablab\cdot\bm{\xi} &= 0.
  \label{eq:div}
\end{align}
We note that the linear equation for $\delta\mathbf{B}$ (\ref{eq:linear_ideal}) can be recovered from \eq(\ref{eq:ideal_xi})
using the relation (\ref{eq:deltaB}).
Finally, we note that the perturbation to the Lorentz force is
\begin{align}
  \delta(\mathbf{J}\times\mathbf{B}) &= \delta\mathbf{J}\times\mathbf{B} + \mathbf{J}\times\delta\mathbf{B} \nonumber \\
  &= \mathcal{F}(\bm{\xi}),
\end{align}
where $\mathcal{F}$ is the linear operator
\begin{align}
  \mathcal{F}(\bm{\xi}) &= \mathbf{B}\times\nablab\times\nablab\times\mathbf{B}\times\bm{\xi}
    - \mathbf{J}\times\nablab\times\mathbf{B}\times\bm{\xi}. \label{eq:F}
\end{align}
Significantly, the operator $\mathcal{F}$ is self-adjoint over the space of (complex) vector fields satisfying the constraint
(\ref{eq:div}) and with respect to the inner product $\langle\cdot,\cdot\rangle$ defined as
\begin{equation}
  \langle\bm{\alpha},\bm{\beta}\rangle \equiv \int\!\dd V\,\bm{\alpha}^\star\cdot\bm{\beta},
  \label{eq:brackets}
\end{equation}
where $\bm{\alpha}^\star$ denotes the complex conjugate of $\bm{\alpha}$.
To prove the self-adjointness of $\mathcal{F}$, we make use of the identity
\begin{align}
  \nablab(\bm{\alpha}\cdot\bm{\beta}\times\bm{\gamma})
  + (\bm{\alpha}\times\bm{\beta})\nablab\cdot\bm{\gamma}
  + (\bm{\beta}\times\bm{\gamma})\nablab\cdot\bm{\alpha}
  + (\bm{\gamma}\times\bm{\alpha})\nablab\cdot\bm{\beta}
  = \nonumber \\
  \bm{\alpha}\times\left(\nablab\times(\bm{\beta}\times\bm{\gamma})\right)
  + \bm{\beta}\times\left(\nablab\times(\bm{\gamma}\times\bm{\alpha})\right)
  + \bm{\gamma}\times\left(\nablab\times(\bm{\alpha}\times\bm{\beta})\right),
\end{align}
which holds for any triple $\bm{\alpha}, \bm{\beta}, \bm{\gamma}$.
If the background field $\mathbf{B}$ satisfies
the equilibrium condition (\ref{eq:ideal_equilibrium}),
then it follows that
\begin{equation}
  \nablab(\bm{\alpha}\cdot\mathbf{B}\times\mathbf{J})
  + (\mathbf{B}\times\mathbf{J})\nablab\cdot\bm{\alpha}
  =
  \mathbf{B}\times\left(\nablab\times(\mathbf{J}\times\bm{\alpha})\right)
  + \mathbf{J}\times\left(\nablab\times(\bm{\alpha}\times\mathbf{B})\right).
\end{equation}
From this result,
and using integration by parts repeatedly, we find that
\begin{align}
  \langle\bm{\alpha},\mathcal{F}(\bm{\beta})\rangle - \langle\mathcal{F}(\bm{\alpha}),\bm{\beta}\rangle
  =
  \int\!\dd V\,\nablab\cdot[&
    (\mathbf{B}\times\bm{\alpha}^\star)\times(\nablab\times\mathbf{B}\times\bm{\beta}) \nonumber \\
    &- (\mathbf{B}\times\bm{\beta})\times(\nablab\times\mathbf{B}\times\bm{\alpha}^\star)
    - \mathbf{J}\cdot(\bm{\alpha}^\star\times\bm{\beta})\mathbf{B}
  ]
\end{align}
for any two divergence-free vector fields $\bm{\alpha}$ and $\bm{\beta}$.
Therefore if $\bm{\xi}_1$ and $\bm{\xi}_2$ are two solutions of \eqs(\ref{eq:ideal_xi})--(\ref{eq:div}),
and if boundary conditions are chosen appropriately, then
\begin{equation}
  \langle\bm{\xi}_1,\mathcal{F}(\bm{\xi}_2)\rangle = \langle\mathcal{F}(\bm{\xi}_1),\bm{\xi}_2\rangle,
\end{equation}
demonstrating that $\mathcal{F}$ is indeed self-adjoint.
It can also be shown that
\begin{align}
  \frac{\dd}{\dd t}\langle\bm{\xi}_1,\mathcal{F}(\bm{\xi}_2)\rangle =
  \int\!\dd V\,\nablab\cdot&\left[
    (\mathbf{B}\times\bm{\xi}_1^\star)\times\nablab\times\mathbf{B}\times\frac{\partial\bm{\xi}_2}{\partial t}
    + (\nablab\times\mathbf{B}\times\bm{\xi}_1^\star)\times\mathbf{B}\times\frac{\partial\bm{\xi}_2}{\partial t}
    \right.\nonumber \\
    &\hspace{3cm}\left.- (\mathbf{J}\times\bm{\xi}_1^\star)\times\mathbf{B}\times\frac{\partial\bm{\xi}_2}{\partial t}
    -(\mathbf{J}\times\mathbf{B}\cdot\bm{\xi}_2)\frac{\partial\bm{\xi}_1^\star}{\partial t}
  \right],
\end{align}
so the quantity $\langle\bm{\xi}_1,\mathcal{F}(\bm{\xi}_2)\rangle$ is conserved for suitable boundary conditions.
This implies, in particular, that if $\bm{\xi}$ is an eigensolution of \eq(\ref{eq:ideal_xi}) with growth-rate $\lambda$
then
\begin{equation}
    \frac{\dd}{\dd t}\langle\bm{\xi},\mathcal{F}(\bm{\xi})\rangle =
    (\lambda+\lambda^\star)\langle\bm{\xi},\mathcal{F}(\bm{\xi})\rangle = 0.
\end{equation}
Therefore any unstable mode ($\lambda+\lambda^\star > 0$)
must be a solution of the equation $\langle\bm{\xi},\mathcal{F}(\bm{\xi})\rangle=0$.
Since $\mathcal{F}$ is a self-adjoint operator, a necessary condition for instability is therefore that $\mathcal{F}$
has both positive and negative eigenvalues.
Furthermore, it can be shown from \eq(\ref{eq:F}) that
\begin{align}
  \mathcal{F}(\bm{\xi}) &= (\mathbf{B}\cdot\nablab)^2\bm{\xi} - \nablab(\mathbf{B}\cdot\delta\mathbf{B})
  - \bm{\xi}\cdot\nablab(\mathbf{B}\cdot\nablab\mathbf{B})
  - (\nablab\cdot\bm{\xi})\mathbf{B}\cdot\nablab\mathbf{B}
  - \mathbf{B}\cdot\nablab[(\nablab\cdot\bm{\xi})\mathbf{B}].
\end{align}
From this, and using the fact that any EMHD equilibrium must have
\begin{equation}
  \mathbf{B}\cdot\nablab\mathbf{B} = \nablab\psi
\end{equation}
for some function $\psi$, it follows that
\begin{equation}
  \langle\bm{\xi},\mathcal{F}(\bm{\xi})\rangle
  =
  -\int\!\dd V\,\left[
    |\mathbf{B}\cdot\nablab\bm{\xi}|^2
    + \xi_i\frac{\partial^2\psi}{\partial x_i\partial x_j}\xi^\star_j
  \right].
  \label{eq:Lundquist}
\end{equation}
We deduce immediately the result mentioned in \S\ref{sec:straight},
that a field with $\mathbf{B}\cdot\nablab\mathbf{B}=0$ is always stable,
since then the right-hand side of \eq(\ref{eq:Lundquist}) is always negative.

The quantity in \eq(\ref{eq:Lundquist}) arises in another context ---
that of ``regular'' MHD for an incompressible fluid \cite{Lundquist51}.
In that context, the necessary \emph{and sufficient} condition for instability of a static equilibrium
is that $\langle\bm{\xi},\mathcal{F}(\bm{\xi})\rangle$ is positive for some perturbation $\bm{\xi}$.
This is because $\langle\bm{\xi},\mathcal{F}(\bm{\xi})\rangle$ represents
the change in magnetic energy produced by the perturbation (to second order in $\bm{\xi}$),
so a positive eigenvalue of the operator $\mathcal{F}$ represents a perturbation that converts energy
stored in the \Bf\ into kinetic energy.
In EMHD, on the other hand, the total magnetic energy is always conserved,
and the inertia-less electron fluid has no kinetic energy.
The only perturbations that can grow, therefore, are those that do not change the total magnetic energy of the system,
\ie\ those that have $\langle\bm{\xi},\mathcal{F}(\bm{\xi})\rangle = 0$.

In summary, the necessary and sufficient condition for instability of a static \Bf\ in incompressible MHD
is a necessary condition for instability in EMHD.  This means that EMHD is ``at least as stable'' as incompressible MHD,
because all field configurations that are stable in MHD are also stable in EMHD.
In fact, we can say that EMHD is \emph{more} stable than incompressible MHD, because there are fields that are unstable in MHD but stable in EMHD.  For example, a purely toroidal field of the form
$\mathbf{B} = \mathbf{e}_z\times\mathbf{x}$ is subject to ``kink'' instability in MHD \cite{Tayler57},
but is stable in EMHD, as can be deduced immediately from \eq(\ref{eq:shear}).

\section{Non-uniform electron density}
\label{sec:non-uniform}

In a neutron star crust the electron density varies by several orders of magnitude \cite{ChamelHaensel08},
with a typical density scale-height of $10^3$--$10^4\cm$.
Therefore, the consequences of density gradients for EMHD stability must
be considered in any realistic model of neutron star fields.
In this section we extend our previous results to take account of such gradients.
To simplify the analysis we will neglect resistivity from here on, and so
our basic equation is
\begin{equation}
  \frac{\partial\mathbf{B}}{\partial t}
  =
  \nablab\times\left[
    \frac{1}{n}\mathbf{B}\times(\nablab\times\mathbf{B})
  \right],
\end{equation}
where the electron density $n(\mathbf{x})$ is prescribed,
and we consider linear perturbations $\delta\mathbf{B}(\mathbf{x},t)$
to a steady equilibrium solution $\mathbf{B}(\mathbf{x})$.

\subsection{\Dsi}
We begin with the simplest example of an EMHD equilibrium field with non-uniform density,
which is $\mathbf{B} = B(z)\,\mathbf{e}_x$ and $n = n(z)$.
To avoid any issues arising from singularities and boundary conditions, we will assume from here on that the domain
is infinite, and that $B(z)$ and $n(z)$ are strictly positive for all $z$ and
remain bounded
as $|z| \to \infty$.
As in \S\ref{sec:straight}, we seek \Bf\ perturbations of the form (\ref{eq:ansatz}).
In place of \eq(\ref{eq:2d_linear}) we now find
\begin{equation}
  \frac{b_z''}{b_z} = \frac{(B'/n)'}{B/n} + k_x^2 + k_y^2
    + \left(\frac{\lambda n}{Bk_x} - \ii\frac{B'k_y}{Bk_x}\right)\left(\frac{\lambda n}{Bk_x} - \ii\frac{n'k_y}{nk_x}\right).
    \label{eq:linear3d_full}
\end{equation}
As with \eq(\ref{eq:2d_linear}), a necessary condition for instability is that the first term on the right-hand side is
negative for some range of $z$.
We note that the (dimensionless) electron velocity in the background state is
$\mathbf{v} = -(\nablab\times\mathbf{B})/n = - (B'/n)\mathbf{e}_y$,
so instability requires the presence of electron shear \cite{Cumming-etal04}.
However, as in \S\ref{sec:straight}, this condition is only necessary, and not sufficient.
A more stringent necessary condition can be obtained by the same process that led to \eq(\ref{eq:bound}),
which this time leads to the result
\begin{equation}
  \lambda^2 = \dfrac{\displaystyle k_x^2\int\!\dd z\,\frac{B'n'}{Bn}|b_z|^2 - k_x^2\int\!\dd z\,\left|b_z' - \dfrac{B'}{B}b_z\right|^2
  - k_x^4\int\!\dd z\,|b_z|^2}{\displaystyle\int\!\dd z\,\dfrac{n^2}{B^2}|b_z|^2}
  \label{eq:drive}
\end{equation}
for perturbations with $k_y=0$.
So a necessary condition for instability is that $\dfrac{B'n'}{Bn} > 0$ for some range of $z$.
A similar result was obtained in Ref.~\onlinecite{GordeevRudakov69} for instabilities of a magneto-sonic wavefront.
However, they were only able to demonstrate instability in the limit where the width of the front becomes
much narrower than the wavelength of the perturbations.
In fact, instabilities can be found under much more general conditions.
For example, suppose that $B(z) = n(z) = \sech^\gamma(z)$ for some positive constant $\gamma$.
Then the general solution
of \eq(\ref{eq:linear3d_full}) can be expressed in terms of Jacobi polynomials:
\begin{equation}
  b_z = \sech^\alpha(z)\exp(-\ii\beta z)P_m^{(\alpha+\ii\beta,\alpha-\ii\beta)}(\tanh(z))
\end{equation}
where the parameters $\alpha$, $\beta$, and $m$ are related to $k_x$, $k_y$, and $\lambda$ by the equations
\begin{align}
  \alpha\beta &= \gamma\lambda k_y/k_x^2 \label{eq:Jacobi1} \\
  \alpha^2-\beta^2 &= k_x^2 + k_y^2 + (\lambda^2-\gamma^2 k_y^2)/k_x^2 \\
  (\alpha+m)(\alpha+m+1) &= \gamma - \gamma^2k_y^2/k_x^2. \label{eq:Jacobi3}
\end{align}
The solution is regular for any positive integer $m$, which we may regard as the (discrete) vertical wavenumber.
For given $k_x$, $k_y$, and $m$, equations (\ref{eq:Jacobi1})--(\ref{eq:Jacobi3})
implicitly provide a dispersion relation for $\lambda$.
In general, perturbations with large wavenumbers $k_x, k_y$ are purely oscillatory, and perturbations
with sufficiently small $k_x, k_y$ are either stable or unstable.
The fastest growing unstable mode has $m=\beta=k_y=0$, and is of the form
\begin{equation}
  \delta\mathbf{B} \propto \left(\begin{array}{c}
    \mp\ii\sqrt{2}\tanh(z) \\
    1 \\
    1
  \end{array}\right)\sech^\alpha(z)\exp(\tfrac{\alpha^2}{2}t \pm \ii\tfrac{\alpha}{\sqrt{2}}x),
  \label{eq:eigenmode}
\end{equation}
where $\alpha(\alpha+1) = \gamma$.
We note that the fastest growing mode has $k_y=0$ (\ie\ $\mathbf{J}\cdot\mathbf{k} = 0$)
and is therefore non-oscillatory, in accordance with \eq(\ref{eq:drive}).

\subsubsection{Limiting cases}

With the choice of profiles $B(z) = n(z) = \sech^\gamma(z)$ used above,
the electron velocity profile is $v = -B'/n = \gamma\tanh(z)$,
and so the electron shear profile, $v' = \gamma\,\sech^2(z)$,
is either wider or narrower than the $B(z)$ and $n(z)$ profiles,
depending on whether $\gamma>2$ or $\gamma<2$.
The fastest-growing eigenmode (\ref{eq:eigenmode})
has a characteristic lengthscale $\simeq 1/\alpha$ in the $x$ and $z$ directions, which
is always at least as wide as the $B(z)$ and $n(z)$ profiles,
but can be wider or narrower than the $v'(z)$ profile, depending on $\gamma$.

By taking the asymptotic limits $\gamma \to 0$ and $\gamma \to \infty$ we obtain two interesting limiting cases.
More precisely, noting that $\sech^\gamma(z/\gamma) \to \exp(-|z|)$ as $\gamma \to 0$,
we deduce from \eq(\ref{eq:eigenmode}) that the fastest growing mode for the background with $B = n = \exp(-|z|)$ has
$\delta B_z = \exp(-|z|)\exp(\tfrac{1}{2}t \pm \ii\tfrac{1}{\sqrt{2}}x)$.
In this limit the electron velocity becomes a Heaviside function of $z$,
but the lengthscale of the unstable mode remains finite.
Conversely, using $\sech^\gamma(z/\gamma^{1/2}) \to \exp(-z^2\!/2)$ as $\gamma \to \infty$,
we deduce that the fastest growing mode for the background with $B = n = \exp(-z^2\!/2)$ has
$\delta B_z = \exp(\tfrac{1}{2}t \pm \ii\tfrac{1}{\sqrt{2}}x)$.
In this case, the unstable mode has finite amplitude throughout the domain,
even where $B(z)$ and $n(z)$ are exponentially small.
This case is somewhat pathological, however,
because the electron velocity, $v=z$, is unbounded,
and the unstable modes form a continuous spectrum.

\subsubsection{The nature of the instability}
\begin{figure}[h!]
  \centering%
  \includegraphics[width=.5\textwidth]{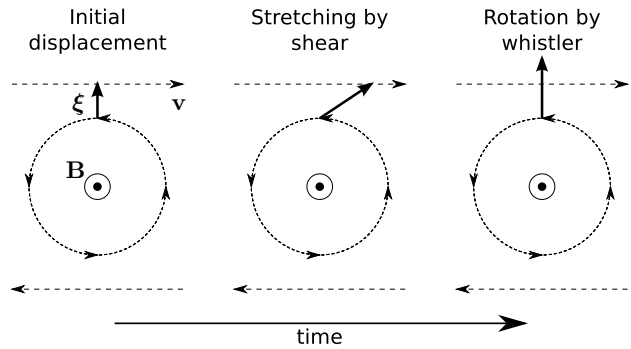}%
  \caption{Cartoon of the \dsi.
    The displacement vector $\bm{\xi}$
    is stretched by the electron velocity $\mathbf{v}$
    and rotated about the direction of the \Bf\ $\mathbf{B}$.
    As a result the initial perturbation is amplified.}%
  \label{fig:schematic}
\end{figure}

It is tempting to think of the \dsi\ as a trapped whistler mode that is advected and sheared
by the electron flow, as described by \eq(\ref{eq:linear}).
Whistler waves are right-hand polarized, so the displacement vector, $\bm{\xi}$, rotates in a sense
that depends on the direction of the \Bf.  If the electron shear acts in opposition to this rotation,
as illustrated in \fig\ref{fig:schematic}, then the perturbation will be locally amplified.
Mathematically, this corresponds to the first term on the right-hand side of \eq(\ref{eq:linear3d_full})
having negative sign.
This is essentially the physical mechanism proposed by %
\citetalias{RheinhardtGeppert02} %
for their Hall drift instability %
(see also Refs.~\onlinecite{Rheinhardt-etal04,Cumming-etal04}). %

However, this physical picture of the instability is incomplete, because it makes no mention
of the electron density gradient, which we have previously shown is necessary for instability.
(Nor can it explain the resistive tearing instability described in \S\ref{sec:tearing}, %
because it makes no mention of either resistivity or the magnetic null surface.) %
The role of the density gradient can be better understood by analogy with magneto-buoyancy instability in regular MHD.
To that end, suppose we have a layer of \Bf\ $\mathbf{B} = B(z)\,\mathbf{e}_x$ in an MHD fluid with density $n = n(z)$
and a gravitational potential $\phi = \phi(z)$.  If we seek marginally stable ($\lambda=0$) perturbations
of the form given by \eq(\ref{eq:ansatz}), then we eventually arrive at an equation \cite{HughesCattaneo87}
\begin{equation}
  \frac{b_z''}{b_z} = \frac{B''}{B} + \left(1 + \frac{k_y^2}{k_x^2}\right)
    \left[k_x^2 - \frac{\phi'}{|\mathbf{v}_{\!\rm A}|^2}\left(\frac{n'}{n}+\frac{\phi'}{a^2}\right)\right],
    \label{eq:HC87}
\end{equation}
where $\mathbf{v}_{\!\rm A}$ is the Alfv\'en velocity, and $a$ is the sound speed.
Assuming that the background state is an adiabatic, hydrostatic balance between fluid pressure, magnetic pressure,
and gravity, it follows that
\begin{equation}
  \frac{n'}{n}+\frac{\phi'}{a^2} = -\frac{|\mathbf{v}_{\!\rm A}|^2}{a^2}\frac{B'}{B},
\end{equation}
and so \eq(\ref{eq:HC87}) can equivalently be written as
\begin{equation}
  \frac{b_z''}{b_z} = \frac{B''}{B} + \left(1 + \frac{k_y^2}{k_x^2}\right)
    \left[k_x^2 - \frac{B'}{B}\left(\frac{n'}{n}+\frac{|\mathbf{v}_{\!\rm A}|^2}{a^2}\frac{B'}{B}\right)\right].
    \label{eq:HC872}
\end{equation}
The necessary and sufficient condition for magneto-buoyancy instability is that this equation has bounded solutions.
Taking the limit $k_y \to \infty$, %
bounded solutions can always be obtained if the expression in square brackets is negative for some range of $z$. %
There may also be bounded solutions %
in the opposite limit, $k_y \to 0$, %
but the instability criterion is more stringent in that case. %
In this second limit, \eq(\ref{eq:HC872}) becomes %
almost identical to the \dsi\ equation (\ref{eq:linear3d_full})
with $\lambda = k_y = 0$,
except that it contains an additional term involving the ratio $|\mathbf{v}_{\!\rm A}|/a$,
which accounts for the expansion of rising fluid parcels.  Such a term does not arise in the \dsi\
because the fluid motions are subject to the constraint given by \eq(\ref{eq:div_nxi}),
and so the effective sound speed in EMHD is infinite.
If such a constraint were imposed in the magneto-buoyancy problem, then the stability criterion
for modes with $k_y=0$ %
would exactly match that for \dsi.
Indeed, in MHD this constraint is called the ``anelastic approximation'' \cite{BraginskyRoberts95},
and so the \dsi\ in EMHD is closely analogous to the magneto-buoyancy instability in an anelastic fluid.
This connection between EMHD and anelastic MHD
is the natural generalization of the results presented in \S\ref{sec:Lundquist},
as we demonstrate in the next section.

\subsection{Connection with anelastic MHD}
\label{sec:anelastic}

The (diffusionless) anelastic equations for a barotropic fluid are \cite{BraginskyRoberts95}
\begin{align}
  \frac{\partial\mathbf{v}}{\partial t} + \mathbf{v}\cdot\nablab\mathbf{v} &= \mathbf{J}\times\mathbf{B}/n
    - \nablab(P/n + \phi)\\
  \nablab\cdot(n\mathbf{v}) &= 0 \\
  \frac{\partial\mathbf{B}}{\partial t} &= \nablab\times(\mathbf{v}\times\mathbf{B}),
\end{align}
where $\mathbf{v}$ is now the fluid velocity, $P$ is the fluid pressure, $\phi$ is the gravitational potential,
and $n(\mathbf{x})$ is the fluid density,
which is a prescribed function of position in the anelastic approximation.
We also make Cowling's approximation, in which $\phi$ is taken to be a fixed function of position.
Suppose we have a steady, static equilibrium with $\mathbf{v}=0$ and
\begin{align}
  0 &= \mathbf{J}\times\mathbf{B}/n - \nablab(P/n + \phi) \nonumber \\
  \Rightarrow 0&= \nablab\times\left(\mathbf{J}\times\mathbf{B}/n\right).
  \label{eq:anelastic_equilibrium}
\end{align}
Linear perturbations to this state, when expressed in terms of the Lagrangian fluid perturbation $\bm{\xi}$,
obey the equations
\begin{align}
  n\frac{\partial^2\bm{\xi}}{\partial t^2} &= \mathcal{F}(\bm{\xi}) - n\nablab(\delta P/n) \\
  \nablab\cdot(n\bm{\xi}) &= 0 \label{eq:anelastic},  
\end{align}
where $\mathcal{F}$ is the linear operator defined in \eq(\ref{eq:F}).
As in \S\ref{sec:Lundquist}, it can be shown that $\mathcal{F}$ is a self-adjoint operator over
the space of vector fields satisfying \eq(\ref{eq:anelastic}) with respect to the inner product defined in \eq(\ref{eq:brackets}).
It follows by an energy argument similar to that of Ref.~\onlinecite{Lundquist51} that
the necessary and sufficient condition for instability in this system is that there exists a perturbation $\bm{\xi}$
that satisfies \eq(\ref{eq:anelastic}) and has $\langle\bm{\xi},\mathcal{F}(\bm{\xi})\rangle > 0$.

We now return to the EMHD equations (\ref{eq:xi}) and (\ref{eq:div_nxi}).
We first note that the anelastic equilibrium condition (\ref{eq:anelastic_equilibrium}) is
also the EMHD equilibrium condition, provided that we now interpret $n$ as the electron density,
and $\phi$ as the electric potential.
In this EMHD equilibrium, the Lorentz force is balanced by a combination of the Coulomb force and electron
pressure, rather than by gravity and fluid pressure.
It can be shown, in the same manner as in \S\ref{sec:Lundquist},
that $\langle\bm{\xi},\mathcal{F}(\bm{\xi})\rangle$ is still a conserved quantity in EMHD with non-uniform density,
and must therefore vanish for any unstable eigenmode.
So the necessary and sufficient condition for instability in anelastic MHD is also (but only)
a necessary condition for instability in EMHD.
Finally, the generalization of \eq(\ref{eq:Lundquist}) in this case is
\begin{equation}
  \langle\bm{\xi},\mathcal{F}(\bm{\xi})\rangle
  =
  -\int\!\dd V\,\left[
    |\mathbf{B}\cdot\nablab\bm{\xi} + (\bm{\xi}\cdot\nablab\ln n)\mathbf{B}|^2
    + n\xi_i\frac{\partial^2\psi}{\partial x_i\partial x_j}\xi^\star_j
    - \tfrac{1}{2}|\mathbf{B}|^2n\xi_i\frac{\partial^2(1/n)}{\partial x_i\partial x_j}\xi^\star_j
  \right],
\end{equation}
where the function $\psi$ is now defined such that
\begin{equation}
  \mathbf{B}\cdot\nablab\mathbf{B} = n\nablab\psi + \tfrac{1}{2}|\mathbf{B}|^2\nablab\ln n.
\end{equation}

This connection between EMHD and anelastic MHD implies that for any instability in the former there must be
a corresponding instability in the latter.
The connection between the EMHD \dsi\ and the anelastic magneto-buoyancy instability is just one such example.

\section{Summary and discussion}
Most known instabilities in EMHD require either finite conductivity or finite electron inertia.
We have demonstrated the existence of a very different instability that, instead, requires electron density gradients
and electron velocity shear.
This \dsi\ may play an important role in the
evolution of \Bfs\ in the crusts of neutron stars,
as well as in nearly-collisionless plasmas on scales smaller than the ion skin depth.
The possibility of such an instability was first recognized in Ref.~\onlinecite{GordeevRudakov69}
in the context of laboratory fusion devices,
and has received very little attention in other contexts.
The instability grows on the Hall timescale,
which is generally faster than the growth-rate of any resistive tearing instability,
such as the so-called Hall drift instability of \citetalias{RheinhardtGeppert02}.
Moreover, the \dsi\ does not require peculiar \Bf\ configurations or boundary conditions
to operate.

It is highly likely that the instabilities observed in
Ref.~\onlinecite{Rheinhardt-etal04},
in a numerical model of a neutron star crust,
include both the resistive tearing instability and the \dsi.
Although they interpreted all of their results in terms of the Hall drift instability,
in fact the instabilities they described clearly form two distinct families.  One family occurs close to
magnetic null surfaces and has a growth-rate that depends on resistivity, as we would expect for a resistive
tearing instability.  The other family occurs in the region where density and magnetic pressure gradients are parallel
and has a growth-rate that is independent of resistivity, as we would expect for the \dsi.

We have demonstrated a connection between the stability properties of EMHD and those of
incompressible/anelastic MHD, which shows the futility in seeking EMHD instabilities for field configurations
that are already known to be stable in MHD.
The \dsi\ is analogous to magneto-buoyancy instability in an anelastic fluid.
Under this analogy, the EMHD fluid, which has zero mass and finite charge,
becomes an anelastic fluid, which has finite mass and zero charge,
and the electric	 potential becomes the gravitational potential.
We have not found any instabilities in ideal, inertia-less EMHD with uniform density;
at present, it is not known whether any such instabilities exist.

By analogy with magneto-buoyancy instability in the solar interior \cite{Parker55-buoyancy}
we suggest that the \dsi\ will greatly enhance the transport of magnetic flux from the
superconducting core of a neutron star to its surface.  This could explain the rapid decrease in the \Bf\
strengths observed in young neutron stars \cite{Lyne-etal85,NarayanOstriker90}.
This might also be the explanation for the magnetic spots suggested by Ref.~\onlinecite{Geppert-etal13}.

Of course, the EMHD equilibrium states and \Bf\ geometries considered in this paper are rather idealized,
and the true situation in neutron star crusts is surely more complex.  A full appreciation of the relevance of these
results to neutron stars can only come from more realistic, direct numerical simulations, such as those of
Refs.~\onlinecite{HollerbachRudiger04,Geppert-etal13,GC14}.

\begin{acknowledgments}
T.S.W.~and R.H.~were supported by STFC grant ST/K000853/1.
M.L.~was supported by NASA grants NNX12AF92G and NNX13AP04G.
We would like to thank Martin Goldman, Serguei Komissarov, Mark Rosin, and Dmitri Uzdensky for helpful discussions.
\end{acknowledgments}


\begin{thebibliography}{50}%
\makeatletter
\providecommand \@ifxundefined [1]{%
 \@ifx{#1\undefined}
}%
\providecommand \@ifnum [1]{%
 \ifnum #1\expandafter \@firstoftwo
 \else \expandafter \@secondoftwo
 \fi
}%
\providecommand \@ifx [1]{%
 \ifx #1\expandafter \@firstoftwo
 \else \expandafter \@secondoftwo
 \fi
}%
\providecommand \natexlab [1]{#1}%
\providecommand \enquote  [1]{``#1''}%
\providecommand \bibnamefont  [1]{#1}%
\providecommand \bibfnamefont [1]{#1}%
\providecommand \citenamefont [1]{#1}%
\providecommand \href@noop [0]{\@secondoftwo}%
\providecommand \href [0]{\begingroup \@sanitize@url \@href}%
\providecommand \@href[1]{\@@startlink{#1}\@@href}%
\providecommand \@@href[1]{\endgroup#1\@@endlink}%
\providecommand \@sanitize@url [0]{\catcode `\\12\catcode `\$12\catcode
  `\&12\catcode `\#12\catcode `\^12\catcode `\_12\catcode `\%12\relax}%
\providecommand \@@startlink[1]{}%
\providecommand \@@endlink[0]{}%
\providecommand \url  [0]{\begingroup\@sanitize@url \@url }%
\providecommand \@url [1]{\endgroup\@href {#1}{\urlprefix }}%
\providecommand \urlprefix  [0]{URL }%
\providecommand \Eprint [0]{\href }%
\providecommand \doibase [0]{http://dx.doi.org/}%
\providecommand \selectlanguage [0]{\@gobble}%
\providecommand \bibinfo  [0]{\@secondoftwo}%
\providecommand \bibfield  [0]{\@secondoftwo}%
\providecommand \translation [1]{[#1]}%
\providecommand \BibitemOpen [0]{}%
\providecommand \bibitemStop [0]{}%
\providecommand \bibitemNoStop [0]{.\EOS\space}%
\providecommand \EOS [0]{\spacefactor3000\relax}%
\providecommand \BibitemShut  [1]{\csname bibitem#1\endcsname}%
\let\auto@bib@innerbib\@empty
\bibitem [{\citenamefont {{Kingsep}}, \citenamefont {{Chukbar}},\ and\
  \citenamefont {{Ian'kov}}(1987)}]{Kingsep-etal87}%
  \BibitemOpen
  \bibfield  {author} {\bibinfo {author} {\bibfnamefont {A.~S.}\ \bibnamefont
  {{Kingsep}}}, \bibinfo {author} {\bibfnamefont {K.~V.}\ \bibnamefont
  {{Chukbar}}},\ and\ \bibinfo {author} {\bibfnamefont {V.~V.}\ \bibnamefont
  {{Ian'kov}}},\ }\href@noop {} {\bibfield  {journal} {\bibinfo  {journal}
  {Voprosy Teorii Plazmy}\ }\textbf {\bibinfo {volume} {16}},\ \bibinfo {pages}
  {209} (\bibinfo {year} {1987})}\BibitemShut {NoStop}%
\bibitem [{\citenamefont {{Gordeev}}, \citenamefont {{Kingsep}},\ and\
  \citenamefont {{Rudakov}}(1994)}]{Gordeev-etal94}%
  \BibitemOpen
  \bibfield  {author} {\bibinfo {author} {\bibfnamefont {A.~V.}\ \bibnamefont
  {{Gordeev}}}, \bibinfo {author} {\bibfnamefont {A.~S.}\ \bibnamefont
  {{Kingsep}}},\ and\ \bibinfo {author} {\bibfnamefont {L.~I.}\ \bibnamefont
  {{Rudakov}}},\ }\href {\doibase 10.1016/0370-1573(94)90097-3} {\bibfield
  {journal} {\bibinfo  {journal} {Phys.~Rep.}\ }\textbf {\bibinfo {volume}
  {243}},\ \bibinfo {pages} {215} (\bibinfo {year} {1994})}\BibitemShut
  {NoStop}%
\bibitem [{\citenamefont {{Gordeev}}\ and\ \citenamefont
  {{Rudakov}}(1969)}]{GordeevRudakov69}%
  \BibitemOpen
  \bibfield  {author} {\bibinfo {author} {\bibfnamefont {A.~V.}\ \bibnamefont
  {{Gordeev}}}\ and\ \bibinfo {author} {\bibfnamefont {L.~I.}\ \bibnamefont
  {{Rudakov}}},\ }\href@noop {} {\bibfield  {journal} {\bibinfo  {journal}
  {J.~Exp.~Theor.~Phys.}\ }\textbf {\bibinfo {volume} {28}},\ \bibinfo {pages}
  {1226} (\bibinfo {year} {1969})}\BibitemShut {NoStop}%
\bibitem [{\citenamefont {{Mandt}}, \citenamefont {{Denton}},\ and\
  \citenamefont {{Drake}}(1994)}]{Mandt-etal94}%
  \BibitemOpen
  \bibfield  {author} {\bibinfo {author} {\bibfnamefont {M.~E.}\ \bibnamefont
  {{Mandt}}}, \bibinfo {author} {\bibfnamefont {R.~E.}\ \bibnamefont
  {{Denton}}},\ and\ \bibinfo {author} {\bibfnamefont {J.~F.}\ \bibnamefont
  {{Drake}}},\ }\href {\doibase 10.1029/93GL03382} {\bibfield  {journal}
  {\bibinfo  {journal} {Geophys.~Res.~Lett.}\ }\textbf {\bibinfo {volume}
  {21}},\ \bibinfo {pages} {73} (\bibinfo {year} {1994})}\BibitemShut {NoStop}%
\bibitem [{\citenamefont {{Avinash}}\ \emph {et~al.}(1998)\citenamefont
  {{Avinash}}, \citenamefont {{Bulanov}}, \citenamefont {{Esirkepov}},
  \citenamefont {{Kaw}}, \citenamefont {{Pegoraro}}, \citenamefont
  {{Sasorov}},\ and\ \citenamefont {{Sen}}}]{Avinash-etal98}%
  \BibitemOpen
  \bibfield  {author} {\bibinfo {author} {\bibfnamefont {K.}~\bibnamefont
  {{Avinash}}}, \bibinfo {author} {\bibfnamefont {S.~V.}\ \bibnamefont
  {{Bulanov}}}, \bibinfo {author} {\bibfnamefont {T.}~\bibnamefont
  {{Esirkepov}}}, \bibinfo {author} {\bibfnamefont {P.}~\bibnamefont {{Kaw}}},
  \bibinfo {author} {\bibfnamefont {F.}~\bibnamefont {{Pegoraro}}}, \bibinfo
  {author} {\bibfnamefont {P.~V.}\ \bibnamefont {{Sasorov}}},\ and\ \bibinfo
  {author} {\bibfnamefont {A.}~\bibnamefont {{Sen}}},\ }\href {\doibase
  10.1063/1.873005} {\bibfield  {journal} {\bibinfo  {journal} {Phys.~Plasmas}\
  }\textbf {\bibinfo {volume} {5}},\ \bibinfo {pages} {2849} (\bibinfo {year}
  {1998})}\BibitemShut {NoStop}%
\bibitem [{\citenamefont {{Deng}}\ and\ \citenamefont
  {{Matsumoto}}(2001)}]{DengMatsumoto01}%
  \BibitemOpen
  \bibfield  {author} {\bibinfo {author} {\bibfnamefont {X.~H.}\ \bibnamefont
  {{Deng}}}\ and\ \bibinfo {author} {\bibfnamefont {H.}~\bibnamefont
  {{Matsumoto}}},\ }\href {\doibase 10.1038/410557A0} {\bibfield  {journal}
  {\bibinfo  {journal} {Nature}\ }\textbf {\bibinfo {volume} {410}},\ \bibinfo
  {pages} {557} (\bibinfo {year} {2001})}\BibitemShut {NoStop}%
\bibitem [{\citenamefont {{Goldreich}}\ and\ \citenamefont
  {{Reisenegger}}(1992)}]{GoldreichReisenegger92}%
  \BibitemOpen
  \bibfield  {author} {\bibinfo {author} {\bibfnamefont {P.}~\bibnamefont
  {{Goldreich}}}\ and\ \bibinfo {author} {\bibfnamefont {A.}~\bibnamefont
  {{Reisenegger}}},\ }\href {\doibase 10.1086/171646} {\bibfield  {journal}
  {\bibinfo  {journal} {Astrophys.~J.}\ }\textbf {\bibinfo {volume} {395}},\
  \bibinfo {pages} {250} (\bibinfo {year} {1992})}\BibitemShut {NoStop}%
\bibitem [{\citenamefont {{Cumming}}, \citenamefont {{Arras}},\ and\
  \citenamefont {{Zweibel}}(2004)}]{Cumming-etal04}%
  \BibitemOpen
  \bibfield  {author} {\bibinfo {author} {\bibfnamefont {A.}~\bibnamefont
  {{Cumming}}}, \bibinfo {author} {\bibfnamefont {P.}~\bibnamefont {{Arras}}},
  \ and\ \bibinfo {author} {\bibfnamefont {E.}~\bibnamefont {{Zweibel}}},\
  }\href {\doibase 10.1086/421324} {\bibfield  {journal} {\bibinfo  {journal}
  {Astrophys.~J.}\ }\textbf {\bibinfo {volume} {609}},\ \bibinfo {pages} {999}
  (\bibinfo {year} {2004})}\BibitemShut {NoStop}%
\bibitem [{\citenamefont {{Pacini}}(1967)}]{Pacini67}%
  \BibitemOpen
  \bibfield  {author} {\bibinfo {author} {\bibfnamefont {F.}~\bibnamefont
  {{Pacini}}},\ }\href {\doibase 10.1038/216567a0} {\bibfield  {journal}
  {\bibinfo  {journal} {Nature}\ }\textbf {\bibinfo {volume} {216}},\ \bibinfo
  {pages} {567} (\bibinfo {year} {1967})}\BibitemShut {NoStop}%
\bibitem [{\citenamefont {{Gold}}(1969)}]{Gold69}%
  \BibitemOpen
  \bibfield  {author} {\bibinfo {author} {\bibfnamefont {T.}~\bibnamefont
  {{Gold}}},\ }\href {\doibase 10.1038/221025a0} {\bibfield  {journal}
  {\bibinfo  {journal} {Nature}\ }\textbf {\bibinfo {volume} {221}},\ \bibinfo
  {pages} {25} (\bibinfo {year} {1969})}\BibitemShut {NoStop}%
\bibitem [{\citenamefont {{Thompson}}\ and\ \citenamefont
  {{Duncan}}(1995)}]{ThompsonDuncan95}%
  \BibitemOpen
  \bibfield  {author} {\bibinfo {author} {\bibfnamefont {C.}~\bibnamefont
  {{Thompson}}}\ and\ \bibinfo {author} {\bibfnamefont {R.~C.}\ \bibnamefont
  {{Duncan}}},\ }\href@noop {} {\bibfield  {journal} {\bibinfo  {journal}
  {Mon.~Not.~R.~Astron.~Soc.}\ }\textbf {\bibinfo {volume} {275}},\ \bibinfo
  {pages} {255} (\bibinfo {year} {1995})}\BibitemShut {NoStop}%
\bibitem [{\citenamefont {{Thompson}}\ and\ \citenamefont
  {{Duncan}}(1996)}]{ThompsonDuncan96}%
  \BibitemOpen
  \bibfield  {author} {\bibinfo {author} {\bibfnamefont {C.}~\bibnamefont
  {{Thompson}}}\ and\ \bibinfo {author} {\bibfnamefont {R.~C.}\ \bibnamefont
  {{Duncan}}},\ }\href {\doibase 10.1086/178147} {\bibfield  {journal}
  {\bibinfo  {journal} {Astrophys.~J.}\ }\textbf {\bibinfo {volume} {473}},\
  \bibinfo {pages} {322} (\bibinfo {year} {1996})}\BibitemShut {NoStop}%
\bibitem [{\citenamefont {{Levin}}\ and\ \citenamefont
  {{Lyutikov}}(2012)}]{LevinLyutikov12}%
  \BibitemOpen
  \bibfield  {author} {\bibinfo {author} {\bibfnamefont {Y.}~\bibnamefont
  {{Levin}}}\ and\ \bibinfo {author} {\bibfnamefont {M.}~\bibnamefont
  {{Lyutikov}}},\ }\href {\doibase 10.1111/j.1365-2966.2012.22016.x} {\bibfield
   {journal} {\bibinfo  {journal} {Mon.~Not.~R.~Astron.~Soc.}\ }\textbf
  {\bibinfo {volume} {427}},\ \bibinfo {pages} {1574} (\bibinfo {year}
  {2012})}\BibitemShut {NoStop}%
\bibitem [{\citenamefont {{Lyutikov}}(2003)}]{Lyutikov03}%
  \BibitemOpen
  \bibfield  {author} {\bibinfo {author} {\bibfnamefont {M.}~\bibnamefont
  {{Lyutikov}}},\ }\href {\doibase 10.1046/j.1365-2966.2003.07110.x} {\bibfield
   {journal} {\bibinfo  {journal} {Mon.~Not.~R.~Astron.~Soc.}\ }\textbf
  {\bibinfo {volume} {346}},\ \bibinfo {pages} {540} (\bibinfo {year}
  {2003})}\BibitemShut {NoStop}%
\bibitem [{\citenamefont {{Lyutikov}}(2006)}]{Lyutikov06}%
  \BibitemOpen
  \bibfield  {author} {\bibinfo {author} {\bibfnamefont {M.}~\bibnamefont
  {{Lyutikov}}},\ }\href {\doibase 10.1111/j.1365-2966.2006.10069.x} {\bibfield
   {journal} {\bibinfo  {journal} {Mon.~Not.~R.~Astron.~Soc.}\ }\textbf
  {\bibinfo {volume} {367}},\ \bibinfo {pages} {1594} (\bibinfo {year}
  {2006})}\BibitemShut {NoStop}%
\bibitem [{\citenamefont {{Gordeev}}(1970)}]{Gordeev70}%
  \BibitemOpen
  \bibfield  {author} {\bibinfo {author} {\bibfnamefont {A.~V.}\ \bibnamefont
  {{Gordeev}}},\ }\href {http://stacks.iop.org/0029-5515/10/i=3/a=012}
  {\bibfield  {journal} {\bibinfo  {journal} {Nucl.~Fusion}\ }\textbf {\bibinfo
  {volume} {10}},\ \bibinfo {pages} {319} (\bibinfo {year} {1970})}\BibitemShut
  {NoStop}%
\bibitem [{\citenamefont {{Bulanov}}, \citenamefont {{Pegoraro}},\ and\
  \citenamefont {{Sakharov}}(1992)}]{Bulanov-etal92}%
  \BibitemOpen
  \bibfield  {author} {\bibinfo {author} {\bibfnamefont {S.~V.}\ \bibnamefont
  {{Bulanov}}}, \bibinfo {author} {\bibfnamefont {F.}~\bibnamefont
  {{Pegoraro}}},\ and\ \bibinfo {author} {\bibfnamefont {A.~S.}\ \bibnamefont
  {{Sakharov}}},\ }\href {\doibase 10.1063/1.860467} {\bibfield  {journal}
  {\bibinfo  {journal} {Phys.~Fluids B}\ }\textbf {\bibinfo {volume} {4}},\
  \bibinfo {pages} {2499} (\bibinfo {year} {1992})}\BibitemShut {NoStop}%
\bibitem [{\citenamefont {{Califano}}\ \emph {et~al.}(1999)\citenamefont
  {{Califano}}, \citenamefont {{Prandi}}, \citenamefont {{Pegoraro}},\ and\
  \citenamefont {{Bulanov}}}]{Califano-etal99}%
  \BibitemOpen
  \bibfield  {author} {\bibinfo {author} {\bibfnamefont {F.}~\bibnamefont
  {{Califano}}}, \bibinfo {author} {\bibfnamefont {R.}~\bibnamefont
  {{Prandi}}}, \bibinfo {author} {\bibfnamefont {F.}~\bibnamefont
  {{Pegoraro}}},\ and\ \bibinfo {author} {\bibfnamefont {S.~V.}\ \bibnamefont
  {{Bulanov}}},\ }\href {\doibase 10.1063/1.873538} {\bibfield  {journal}
  {\bibinfo  {journal} {Phys.~Plasmas}\ }\textbf {\bibinfo {volume} {6}},\
  \bibinfo {pages} {2332} (\bibinfo {year} {1999})}\BibitemShut {NoStop}%
\bibitem [{\citenamefont {{Gaur}}\ and\ \citenamefont
  {{Das}}(2012)}]{GaurDas12}%
  \BibitemOpen
  \bibfield  {author} {\bibinfo {author} {\bibfnamefont {G.}~\bibnamefont
  {{Gaur}}}\ and\ \bibinfo {author} {\bibfnamefont {A.}~\bibnamefont {{Das}}},\
  }\href {\doibase 10.1063/1.4731728} {\bibfield  {journal} {\bibinfo
  {journal} {Phys.~Plasmas}\ }\textbf {\bibinfo {volume} {19}},\ \bibinfo
  {pages} {072103} (\bibinfo {year} {2012})}\BibitemShut {NoStop}%
\bibitem [{\citenamefont {{Biskamp}}, \citenamefont {{Schwarz}},\ and\
  \citenamefont {{Drake}}(1996)}]{Biskamp-etal96}%
  \BibitemOpen
  \bibfield  {author} {\bibinfo {author} {\bibfnamefont {D.}~\bibnamefont
  {{Biskamp}}}, \bibinfo {author} {\bibfnamefont {E.}~\bibnamefont
  {{Schwarz}}},\ and\ \bibinfo {author} {\bibfnamefont {J.~F.}\ \bibnamefont
  {{Drake}}},\ }\href {\doibase 10.1103/PhysRevLett.76.1264} {\bibfield
  {journal} {\bibinfo  {journal} {Phys.~Rev.~Lett.}\ }\textbf {\bibinfo
  {volume} {76}},\ \bibinfo {pages} {1264} (\bibinfo {year}
  {1996})}\BibitemShut {NoStop}%
\bibitem [{\citenamefont {{Biskamp}}\ \emph {et~al.}(1999)\citenamefont
  {{Biskamp}}, \citenamefont {{Schwarz}}, \citenamefont {{Zeiler}},
  \citenamefont {{Celani}},\ and\ \citenamefont {{Drake}}}]{Biskamp-etal99}%
  \BibitemOpen
  \bibfield  {author} {\bibinfo {author} {\bibfnamefont {D.}~\bibnamefont
  {{Biskamp}}}, \bibinfo {author} {\bibfnamefont {E.}~\bibnamefont
  {{Schwarz}}}, \bibinfo {author} {\bibfnamefont {A.}~\bibnamefont {{Zeiler}}},
  \bibinfo {author} {\bibfnamefont {A.}~\bibnamefont {{Celani}}},\ and\
  \bibinfo {author} {\bibfnamefont {J.~F.}\ \bibnamefont {{Drake}}},\ }\href
  {\doibase 10.1063/1.873312} {\bibfield  {journal} {\bibinfo  {journal}
  {Phys.~Plasmas}\ }\textbf {\bibinfo {volume} {6}},\ \bibinfo {pages} {751}
  (\bibinfo {year} {1999})}\BibitemShut {NoStop}%
\bibitem [{\citenamefont {{Dastgeer}}\ \emph {et~al.}(2000)\citenamefont
  {{Dastgeer}}, \citenamefont {{Das}}, \citenamefont {{Kaw}},\ and\
  \citenamefont {{Diamond}}}]{Dastgeer-etal00}%
  \BibitemOpen
  \bibfield  {author} {\bibinfo {author} {\bibfnamefont {S.}~\bibnamefont
  {{Dastgeer}}}, \bibinfo {author} {\bibfnamefont {A.}~\bibnamefont {{Das}}},
  \bibinfo {author} {\bibfnamefont {P.}~\bibnamefont {{Kaw}}},\ and\ \bibinfo
  {author} {\bibfnamefont {P.~H.}\ \bibnamefont {{Diamond}}},\ }\href {\doibase
  10.1063/1.873843} {\bibfield  {journal} {\bibinfo  {journal} {Phys.~Plasmas}\
  }\textbf {\bibinfo {volume} {7}},\ \bibinfo {pages} {571} (\bibinfo {year}
  {2000})}\BibitemShut {NoStop}%
\bibitem [{\citenamefont {{Cho}}\ and\ \citenamefont
  {{Lazarian}}(2004)}]{ChoLazarian04}%
  \BibitemOpen
  \bibfield  {author} {\bibinfo {author} {\bibfnamefont {J.}~\bibnamefont
  {{Cho}}}\ and\ \bibinfo {author} {\bibfnamefont {A.}~\bibnamefont
  {{Lazarian}}},\ }\href {\doibase 10.1086/425215} {\bibfield  {journal}
  {\bibinfo  {journal} {Astrophys.~J.}\ }\textbf {\bibinfo {volume} {615}},\
  \bibinfo {pages} {L41} (\bibinfo {year} {2004})}\BibitemShut {NoStop}%
\bibitem [{\citenamefont {{Galtier}}(2008)}]{Galtier08b}%
  \BibitemOpen
  \bibfield  {author} {\bibinfo {author} {\bibfnamefont {S.}~\bibnamefont
  {{Galtier}}},\ }\href {\doibase 10.1029/2007JA012821} {\bibfield  {journal}
  {\bibinfo  {journal} {J.~Geophys.~Res.}\ }\textbf {\bibinfo {volume} {113}},\
  \bibinfo {eid} {A01102} (\bibinfo {year} {2008})}\BibitemShut {NoStop}%
\bibitem [{\citenamefont {{Hollerbach}}\ and\ \citenamefont
  {{R{\"u}diger}}(2002)}]{HollerbachRudiger02}%
  \BibitemOpen
  \bibfield  {author} {\bibinfo {author} {\bibfnamefont {R.}~\bibnamefont
  {{Hollerbach}}}\ and\ \bibinfo {author} {\bibfnamefont {G.}~\bibnamefont
  {{R{\"u}diger}}},\ }\href {\doibase 10.1046/j.1365-8711.2002.05905.x}
  {\bibfield  {journal} {\bibinfo  {journal} {Mon.~Not.~R.~Astron.~Soc.}\
  }\textbf {\bibinfo {volume} {337}},\ \bibinfo {pages} {216} (\bibinfo {year}
  {2002})}\BibitemShut {NoStop}%
\bibitem [{\citenamefont {{Hollerbach}}\ and\ \citenamefont
  {{R{\"u}diger}}(2004)}]{HollerbachRudiger04}%
  \BibitemOpen
  \bibfield  {author} {\bibinfo {author} {\bibfnamefont {R.}~\bibnamefont
  {{Hollerbach}}}\ and\ \bibinfo {author} {\bibfnamefont {G.}~\bibnamefont
  {{R{\"u}diger}}},\ }\href {\doibase 10.1111/j.1365-2966.2004.07307.x}
  {\bibfield  {journal} {\bibinfo  {journal} {Mon.~Not.~R.~Astron.~Soc.}\
  }\textbf {\bibinfo {volume} {347}},\ \bibinfo {pages} {1273} (\bibinfo {year}
  {2004})}\BibitemShut {NoStop}%
\bibitem [{\citenamefont {{Pons}}\ and\ \citenamefont
  {{Geppert}}(2007)}]{PonsGeppert07}%
  \BibitemOpen
  \bibfield  {author} {\bibinfo {author} {\bibfnamefont {J.~A.}\ \bibnamefont
  {{Pons}}}\ and\ \bibinfo {author} {\bibfnamefont {U.}~\bibnamefont
  {{Geppert}}},\ }\href {\doibase 10.1051/0004-6361:20077456} {\bibfield
  {journal} {\bibinfo  {journal} {A\&A}\ }\textbf {\bibinfo {volume} {470}},\
  \bibinfo {pages} {303} (\bibinfo {year} {2007})}\BibitemShut {NoStop}%
\bibitem [{\citenamefont {{Wareing}}\ and\ \citenamefont
  {{Hollerbach}}(2009)}]{WareingHollerbach09a}%
  \BibitemOpen
  \bibfield  {author} {\bibinfo {author} {\bibfnamefont {C.~J.}\ \bibnamefont
  {{Wareing}}}\ and\ \bibinfo {author} {\bibfnamefont {R.}~\bibnamefont
  {{Hollerbach}}},\ }\href {\doibase 10.1063/1.3111033} {\bibfield  {journal}
  {\bibinfo  {journal} {Phys.~Plasmas}\ }\textbf {\bibinfo {volume} {16}},\
  \bibinfo {pages} {042307} (\bibinfo {year} {2009})}\BibitemShut {NoStop}%
\bibitem [{\citenamefont {{Wareing}}\ and\ \citenamefont
  {{Hollerbach}}(2010)}]{WareingHollerbach10}%
  \BibitemOpen
  \bibfield  {author} {\bibinfo {author} {\bibfnamefont {C.~J.}\ \bibnamefont
  {{Wareing}}}\ and\ \bibinfo {author} {\bibfnamefont {R.}~\bibnamefont
  {{Hollerbach}}},\ }\href {\doibase 10.1017/S0022377809990158} {\bibfield
  {journal} {\bibinfo  {journal} {J.~Plasma Phys.}\ }\textbf {\bibinfo {volume}
  {76}},\ \bibinfo {pages} {117} (\bibinfo {year} {2010})}\BibitemShut
  {NoStop}%
\bibitem [{\citenamefont {{Gourgouliatos}}\ and\ \citenamefont
  {{Cumming}}(2014)}]{GC14}%
  \BibitemOpen
  \bibfield  {author} {\bibinfo {author} {\bibfnamefont {K.~N.}\ \bibnamefont
  {{Gourgouliatos}}}\ and\ \bibinfo {author} {\bibfnamefont {A.}~\bibnamefont
  {{Cumming}}},\ }\href {\doibase 10.1093/mnras/stt2300} {\bibfield  {journal}
  {\bibinfo  {journal} {Mon.~Not.~R.~Astron.~Soc.}\ }\textbf {\bibinfo {volume}
  {438}},\ \bibinfo {pages} {1618} (\bibinfo {year} {2014})}\BibitemShut
  {NoStop}%
\bibitem [{\citenamefont {{Rheinhardt}}\ and\ \citenamefont
  {{Geppert}}(2002)}]{RheinhardtGeppert02}%
  \BibitemOpen
  \bibfield  {author} {\bibinfo {author} {\bibfnamefont {M.}~\bibnamefont
  {{Rheinhardt}}}\ and\ \bibinfo {author} {\bibfnamefont {U.}~\bibnamefont
  {{Geppert}}},\ }\href {\doibase 10.1103/PhysRevLett.88.101103} {\bibfield
  {journal} {\bibinfo  {journal} {Phys.~Rev.~Lett.}\ }\textbf {\bibinfo
  {volume} {88}},\ \bibinfo {eid} {101103} (\bibinfo {year}
  {2002})}\BibitemShut {NoStop}%
\bibitem [{\citenamefont {{Rheinhardt}}, \citenamefont {{Konenkov}},\ and\
  \citenamefont {{Geppert}}(2004)}]{Rheinhardt-etal04}%
  \BibitemOpen
  \bibfield  {author} {\bibinfo {author} {\bibfnamefont {M.}~\bibnamefont
  {{Rheinhardt}}}, \bibinfo {author} {\bibfnamefont {D.}~\bibnamefont
  {{Konenkov}}},\ and\ \bibinfo {author} {\bibfnamefont {U.}~\bibnamefont
  {{Geppert}}},\ }\href {\doibase 10.1051/0004-6361:20034078} {\bibfield
  {journal} {\bibinfo  {journal} {A\&A}\ }\textbf {\bibinfo {volume} {420}},\
  \bibinfo {pages} {631} (\bibinfo {year} {2004})}\BibitemShut {NoStop}%
\bibitem [{\citenamefont {{Chamel}}\ and\ \citenamefont
  {{Haensel}}(2008)}]{ChamelHaensel08}%
  \BibitemOpen
  \bibfield  {author} {\bibinfo {author} {\bibfnamefont {N.}~\bibnamefont
  {{Chamel}}}\ and\ \bibinfo {author} {\bibfnamefont {P.}~\bibnamefont
  {{Haensel}}},\ }\href {\doibase 10.12942/lrr-2008-10} {\bibfield  {journal}
  {\bibinfo  {journal} {Living Rev.~Relativ.}\ }\textbf {\bibinfo {volume}
  {11}},\ \bibinfo {pages} {10} (\bibinfo {year} {2008})}\BibitemShut {NoStop}%
\bibitem [{\citenamefont {{Boswell}}\ and\ \citenamefont
  {{Chen}}(1997)}]{BoswellChen97}%
  \BibitemOpen
  \bibfield  {author} {\bibinfo {author} {\bibfnamefont {R.~W.}\ \bibnamefont
  {{Boswell}}}\ and\ \bibinfo {author} {\bibfnamefont {F.~F.}\ \bibnamefont
  {{Chen}}},\ }\href {\doibase 10.1109/27.650898} {\bibfield  {journal}
  {\bibinfo  {journal} {IEEE Transactions on Plasma Science}\ }\textbf
  {\bibinfo {volume} {25}},\ \bibinfo {pages} {1229} (\bibinfo {year}
  {1997})}\BibitemShut {NoStop}%
\bibitem [{\citenamefont {{Newell}}(2010)}]{Newell10}%
  \BibitemOpen
  \bibfield  {author} {\bibinfo {author} {\bibfnamefont {P.~T.}\ \bibnamefont
  {{Newell}}},\ }\href {\doibase 10.1038/467927a} {\bibfield  {journal}
  {\bibinfo  {journal} {Nature}\ }\textbf {\bibinfo {volume} {467}},\ \bibinfo
  {pages} {927} (\bibinfo {year} {2010})}\BibitemShut {NoStop}%
\bibitem [{\citenamefont {{Drake}}, \citenamefont {{Kleva}},\ and\
  \citenamefont {{Mandt}}(1994)}]{Drake-etal94}%
  \BibitemOpen
  \bibfield  {author} {\bibinfo {author} {\bibfnamefont {J.~F.}\ \bibnamefont
  {{Drake}}}, \bibinfo {author} {\bibfnamefont {R.~G.}\ \bibnamefont
  {{Kleva}}},\ and\ \bibinfo {author} {\bibfnamefont {M.~E.}\ \bibnamefont
  {{Mandt}}},\ }\href {\doibase 10.1103/PhysRevLett.73.1251} {\bibfield
  {journal} {\bibinfo  {journal} {Phys.~Rev.~Lett.}\ }\textbf {\bibinfo
  {volume} {73}},\ \bibinfo {pages} {1251} (\bibinfo {year}
  {1994})}\BibitemShut {NoStop}%
\bibitem [{\citenamefont {{Furth}}, \citenamefont {{Killeen}},\ and\
  \citenamefont {{Rosenbluth}}(1963)}]{Furth-etal63}%
  \BibitemOpen
  \bibfield  {author} {\bibinfo {author} {\bibfnamefont {H.~P.}\ \bibnamefont
  {{Furth}}}, \bibinfo {author} {\bibfnamefont {J.}~\bibnamefont {{Killeen}}},\
  and\ \bibinfo {author} {\bibfnamefont {M.~N.}\ \bibnamefont
  {{Rosenbluth}}},\ }\href {\doibase 10.1063/1.1706761} {\bibfield  {journal}
  {\bibinfo  {journal} {Phys.~Fluids}\ }\textbf {\bibinfo {volume} {6}},\
  \bibinfo {pages} {459} (\bibinfo {year} {1963})}\BibitemShut {NoStop}%
\bibitem [{\citenamefont {{Pegoraro}}\ and\ \citenamefont
  {{Schep}}(1986)}]{PegoranoSchep86}%
  \BibitemOpen
  \bibfield  {author} {\bibinfo {author} {\bibfnamefont {F.}~\bibnamefont
  {{Pegoraro}}}\ and\ \bibinfo {author} {\bibfnamefont {T.~J.}\ \bibnamefont
  {{Schep}}},\ }\href {\doibase 10.1088/0741-3335/28/4/003} {\bibfield
  {journal} {\bibinfo  {journal} {Plasma Phys.~Control.~Fusion}\ }\textbf
  {\bibinfo {volume} {28}},\ \bibinfo {pages} {647} (\bibinfo {year}
  {1986})}\BibitemShut {NoStop}%
\bibitem [{\citenamefont {{Fruchtman}}\ and\ \citenamefont
  {{Strauss}}(1993)}]{FruchtmanStrauss93}%
  \BibitemOpen
  \bibfield  {author} {\bibinfo {author} {\bibfnamefont {A.}~\bibnamefont
  {{Fruchtman}}}\ and\ \bibinfo {author} {\bibfnamefont {H.~R.}\ \bibnamefont
  {{Strauss}}},\ }\href {\doibase 10.1063/1.860880} {\bibfield  {journal}
  {\bibinfo  {journal} {Phys.~Fluids B}\ }\textbf {\bibinfo {volume} {5}},\
  \bibinfo {pages} {1408} (\bibinfo {year} {1993})}\BibitemShut {NoStop}%
\bibitem [{Note1()}]{Note1}%
  \BibitemOpen
  \bibinfo {note} {In the ``regular'' MHD case considered in Ref.~\protect
  \rev@citealpnum {Furth-etal63} this approximation can be rigorously justified
  by asymptotic analysis. Whether there is such a rigorous justification in the
  EMHD case is unclear, but a more careful study in Ref.~\protect
  \rev@citealpnum {Attico-etal00} suggests that this approximation is
  reasonable.}\BibitemShut {Stop}%
\bibitem [{\citenamefont {{Gourgouliatos}}\ \emph {et~al.}(2013)\citenamefont
  {{Gourgouliatos}}, \citenamefont {{Cumming}}, \citenamefont {{Reisenegger}},
  \citenamefont {{Armaza}}, \citenamefont {{Lyutikov}},\ and\ \citenamefont
  {{Valdivia}}}]{Gourgouliatos-etal13}%
  \BibitemOpen
  \bibfield  {author} {\bibinfo {author} {\bibfnamefont {K.~N.}\ \bibnamefont
  {{Gourgouliatos}}}, \bibinfo {author} {\bibfnamefont {A.}~\bibnamefont
  {{Cumming}}}, \bibinfo {author} {\bibfnamefont {A.}~\bibnamefont
  {{Reisenegger}}}, \bibinfo {author} {\bibfnamefont {C.}~\bibnamefont
  {{Armaza}}}, \bibinfo {author} {\bibfnamefont {M.}~\bibnamefont
  {{Lyutikov}}},\ and\ \bibinfo {author} {\bibfnamefont {J.~A.}\ \bibnamefont
  {{Valdivia}}},\ }\href {\doibase 10.1093/mnras/stt1195} {\bibfield  {journal}
  {\bibinfo  {journal} {Mon.~Not.~R.~Astron.~Soc.}\ }\textbf {\bibinfo {volume}
  {434}},\ \bibinfo {pages} {2480} (\bibinfo {year} {2013})}\BibitemShut
  {NoStop}%
\bibitem [{\citenamefont {{Lundquist}}(1951)}]{Lundquist51}%
  \BibitemOpen
  \bibfield  {author} {\bibinfo {author} {\bibfnamefont {S.}~\bibnamefont
  {{Lundquist}}},\ }\href {\doibase 10.1103/PhysRev.83.307} {\bibfield
  {journal} {\bibinfo  {journal} {Physical Review}\ }\textbf {\bibinfo {volume}
  {83}},\ \bibinfo {pages} {307} (\bibinfo {year} {1951})}\BibitemShut
  {NoStop}%
\bibitem [{\citenamefont {{Tayler}}(1957)}]{Tayler57}%
  \BibitemOpen
  \bibfield  {author} {\bibinfo {author} {\bibfnamefont {R.~J.}\ \bibnamefont
  {{Tayler}}},\ }\href {\doibase 10.1088/0370-1301/70/1/306} {\bibfield
  {journal} {\bibinfo  {journal} {Proc.~Phys.~Soc.~B}\ }\textbf {\bibinfo
  {volume} {70}},\ \bibinfo {pages} {31} (\bibinfo {year} {1957})}\BibitemShut
  {NoStop}%
\bibitem [{\citenamefont {{Hughes}}\ and\ \citenamefont
  {{Cattaneo}}(1987)}]{HughesCattaneo87}%
  \BibitemOpen
  \bibfield  {author} {\bibinfo {author} {\bibfnamefont {D.~W.}\ \bibnamefont
  {{Hughes}}}\ and\ \bibinfo {author} {\bibfnamefont {F.}~\bibnamefont
  {{Cattaneo}}},\ }\href {\doibase 10.1080/03091928708208806} {\bibfield
  {journal} {\bibinfo  {journal} {Geophys. Astrophys. Fluid Dyn.}\ }\textbf
  {\bibinfo {volume} {39}},\ \bibinfo {pages} {65} (\bibinfo {year}
  {1987})}\BibitemShut {NoStop}%
\bibitem [{\citenamefont {{Braginsky}}\ and\ \citenamefont
  {{Roberts}}(1995)}]{BraginskyRoberts95}%
  \BibitemOpen
  \bibfield  {author} {\bibinfo {author} {\bibfnamefont {S.~I.}\ \bibnamefont
  {{Braginsky}}}\ and\ \bibinfo {author} {\bibfnamefont {P.~H.}\ \bibnamefont
  {{Roberts}}},\ }\href {\doibase 10.1080/03091929508228992} {\bibfield
  {journal} {\bibinfo  {journal} {Geophys. Astrophys. Fluid Dyn.}\ }\textbf
  {\bibinfo {volume} {79}},\ \bibinfo {pages} {1} (\bibinfo {year}
  {1995})}\BibitemShut {NoStop}%
\bibitem [{\citenamefont {{Parker}}(1955)}]{Parker55-buoyancy}%
  \BibitemOpen
  \bibfield  {author} {\bibinfo {author} {\bibfnamefont {E.~N.}\ \bibnamefont
  {{Parker}}},\ }\href {\doibase 10.1086/146010} {\bibfield  {journal}
  {\bibinfo  {journal} {Astrophys.~J.}\ }\textbf {\bibinfo {volume} {121}},\
  \bibinfo {pages} {491} (\bibinfo {year} {1955})}\BibitemShut {NoStop}%
\bibitem [{\citenamefont {{Lyne}}, \citenamefont {{Manchester}},\ and\
  \citenamefont {{Taylor}}(1985)}]{Lyne-etal85}%
  \BibitemOpen
  \bibfield  {author} {\bibinfo {author} {\bibfnamefont {A.~G.}\ \bibnamefont
  {{Lyne}}}, \bibinfo {author} {\bibfnamefont {R.~N.}\ \bibnamefont
  {{Manchester}}},\ and\ \bibinfo {author} {\bibfnamefont {J.~H.}\
  \bibnamefont {{Taylor}}},\ }\href@noop {} {\bibfield  {journal} {\bibinfo
  {journal} {Mon.~Not.~R.~Astron.~Soc.}\ }\textbf {\bibinfo {volume} {213}},\
  \bibinfo {pages} {613} (\bibinfo {year} {1985})}\BibitemShut {NoStop}%
\bibitem [{\citenamefont {{Narayan}}\ and\ \citenamefont
  {{Ostriker}}(1990)}]{NarayanOstriker90}%
  \BibitemOpen
  \bibfield  {author} {\bibinfo {author} {\bibfnamefont {R.}~\bibnamefont
  {{Narayan}}}\ and\ \bibinfo {author} {\bibfnamefont {J.~P.}\ \bibnamefont
  {{Ostriker}}},\ }\href {\doibase 10.1086/168529} {\bibfield  {journal}
  {\bibinfo  {journal} {Astrophys.~J.}\ }\textbf {\bibinfo {volume} {352}},\
  \bibinfo {pages} {222} (\bibinfo {year} {1990})}\BibitemShut {NoStop}%
\bibitem [{\citenamefont {{Geppert}}, \citenamefont {{Gil}},\ and\
  \citenamefont {{Melikidze}}(2013)}]{Geppert-etal13}%
  \BibitemOpen
  \bibfield  {author} {\bibinfo {author} {\bibfnamefont {U.}~\bibnamefont
  {{Geppert}}}, \bibinfo {author} {\bibfnamefont {J.}~\bibnamefont {{Gil}}},\
  and\ \bibinfo {author} {\bibfnamefont {G.}~\bibnamefont {{Melikidze}}},\
  }\href {\doibase 10.1093/mnras/stt1527} {\bibfield  {journal} {\bibinfo
  {journal} {Mon.~Not.~R.~Astron.~Soc.}\ }\textbf {\bibinfo {volume} {435}},\
  \bibinfo {pages} {3262} (\bibinfo {year} {2013})}\BibitemShut {NoStop}%
\bibitem [{\citenamefont {{Attico}}, \citenamefont {{Califano}},\ and\
  \citenamefont {{Pegoraro}}(2000)}]{Attico-etal00}%
  \BibitemOpen
  \bibfield  {author} {\bibinfo {author} {\bibfnamefont {N.}~\bibnamefont
  {{Attico}}}, \bibinfo {author} {\bibfnamefont {F.}~\bibnamefont
  {{Califano}}},\ and\ \bibinfo {author} {\bibfnamefont {F.}~\bibnamefont
  {{Pegoraro}}},\ }\href {\doibase 10.1063/1.874076} {\bibfield  {journal}
  {\bibinfo  {journal} {Phys.~Plasmas}\ }\textbf {\bibinfo {volume} {7}},\
  \bibinfo {pages} {2381} (\bibinfo {year} {2000})}\BibitemShut {NoStop}%
\end{thebibliography}
\end{document}